\documentclass[11pt,a4paper]{article}
\usepackage{jheppub}
\usepackage[numbers,sort&compress]{natbib}
\usepackage{xcolor}
\usepackage{amsmath}
\usepackage{amssymb}
\usepackage{amsfonts}
\usepackage{tocloft}
\usepackage{hyperref}
\usepackage{graphicx}
\usepackage{subcaption} % for subfigures
\usepackage{adjustbox}  % for figures
\usepackage{shuffle}
\usepackage[pages=some]{background}
\usepackage{tikz}
\usepackage{feynmp} % for the Feynman diagrams
\usepackage{slashed}
\usepackage{array}
\usepackage{multirow}
\usepackage{float}
\usepackage{makecell} %  Allows line breaks in table cells
\usepackage{siunitx}
\usetikzlibrary{arrows.meta,calc,chains,shapes.geometric}
\tikzstyle{place}=[shape=rectangle,
       font=\small,
       draw=black,fill=faeng_blue!50,thick,text width=19mm,minimum height=1.1cm, align=center]
\tikzstyle{transition}=[shape=rectangle,font=\small,draw=black,thick,fill=faeng_blue!50,text width=25mm,minimum height=0.9cm, align=center]
\definecolor{faeng_blue}{rgb}{0.0, 0.3, 0.6}

\newcommand{\hs}{\hypersetup{pdftitle={Anatar},pdfcreator={Claude Duhr, Pooja Mukherjee, Andres Vasquez},colorlinks=true,linkcolor=[rgb]{0.15,0.35,0.75},citecolor=[rgb]{0.675,0,0.2},urlcolor=[rgb]{0.15,0.35,0.65}}}
\hs
\newcommand{\anatar}{{\sc Anatar}}
\newcommand{\ant}{{\sc Anatar}}
\newcommand{\mathematica}{{\sc Mathematica}}
\newcommand{\qgraf}{{\sc QGraf}}
\newcommand{\kira}{{\sc Kira}}
\newcommand{\litered}{{\sc LiteRed}}
\newcommand{\fermatica}{{\sc Fermatica}}
\newcommand{\form}{{\sc Form}}

\newcommand{\formC}{{\sc FormCalc}}
\newcommand{\mg}{{\sc MadGraph5\_aMC@NLO}}
\newcommand{\feynrules}{{\sc FeynRules}}

\renewcommand{\unit}{1\!\!1}

\newcommand{\exbox}[1]{\begin{center}\fbox{\parbox[c]{14.8cm}{#1}}\end{center}}
\newcommand{\inline}[2]{\begin{tabular}{rl}{\tt \phantom{a}In[#1]:=}#2\end{tabular}}
\newcommand{\outline}[2]{\begin{tabular}{rl}{\tt Out[#1]:=}#2\end{tabular}}
\newcommand{\breakline}{\rule{14.8cm}{0.5pt}}

\definecolor{lightgrey}{RGB}{235,245,255}
\definecolor{lightgreen}{RGB}{165, 250, 187}
\definecolor{lightgreen}{RGB}{241, 255, 235}
\definecolor{lightgreen}{RGB}{235, 255, 242}
\definecolor{lightgreen}{RGB}{235, 255, 249}
\def\mybtab#1\myetab{\begin{tabular}{p{61mm}p{79mm}}#1\end{tabular}}
\def\btab#1\etab{\begin{tabular}{p{55mm}p{85mm}}#1\end{tabular}}
\def\btabx#1\etabx{\begin{tabular}{p{60mm}p{50mm}}#1\end{tabular}}
\def\btaby#1\etaby{\begin{tabular}{p{15mm}p{95mm}}#1\end{tabular}}
\def\bcen{\begin{center}\begin{small}}
\def\ecen{\end{small}\end{center}}
\def\mybgfb#1\myegfb{\bcen\fcolorbox{black}{lightgreen}{\parbox{148mm}{\mybtab#1\myetab}}\ecen}
\def\bgfb#1\egfb{\bcen\fcolorbox{black}{lightgreen}{\parbox{148mm}{\btab#1\etab}}\ecen}
\def\bgfbx#1\egfbx{\bcen\fcolorbox{black}{lightgreen}{\parbox{118mm}{\btabx#1\etabx}}\ecen}
\def\bgfbalign#1\egfbalign{\bcen\fcolorbox{black}{lightgreen}{\parbox{118mm}{\btaby#1\etaby}}\ecen}
\def\pw{\textasciicircum}
\def\beq{\begin{equation}}
\def\eeq{\end{equation}}
\def\bsp#1\esp{\begin{split}#1\end{split}}
\newcommand{\cA}{\mathcal{A}}
\newcommand{\cD}{\mathcal{D}}
\newcommand{\cF}{\mathcal{F}}
\newcommand{\cN}{\mathcal{N}}
\newcommand{\eps}{\epsilon}
\newcommand{\rd}{\textrm{d}}

\backgroundsetup{
scale=1.2,
color=black,
opacity=0.4,
angle=0,
contents={%
\begin{tikzpicture}[remember picture, overlay]
    \node [shift={(5cm, -10cm)}] 
        { \includegraphics[ scale=0.2]{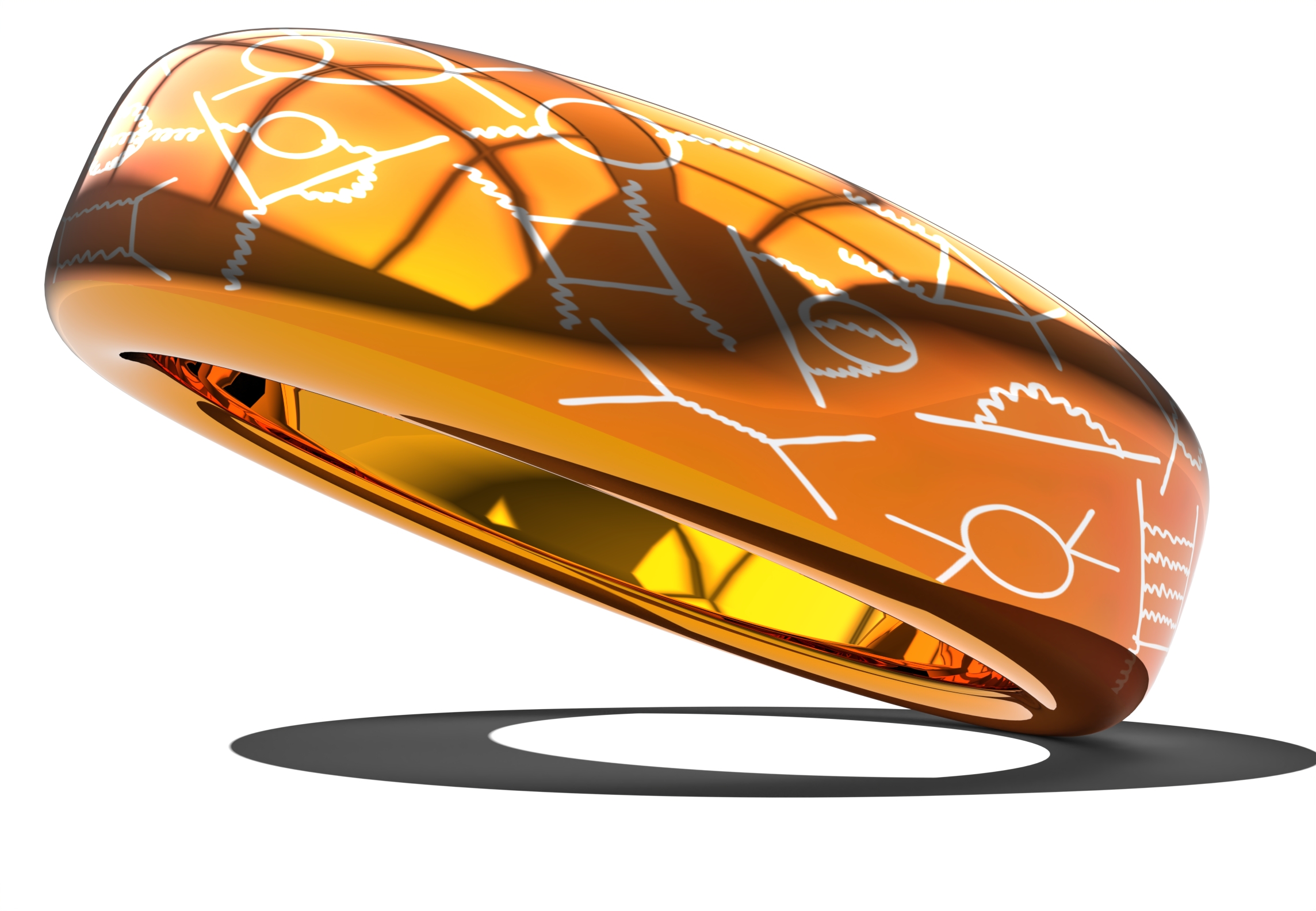}};
\end{tikzpicture}
  }%
}

\title{ANATAR: AN Automated Tool for higher-order Amplitude geneRation}
\author[a]{Claude Duhr}
\author[b]{Pooja Mukherjee}
\author[a]{Andres Vasquez}

\affiliation[a]{Bethe Center for Theoretical Physics, Universit\"at Bonn, D-53115 Bonn, Germany}
\affiliation[b]{\it II. Institut f\"ur Theoretische Physik, Universit\"at Hamburg, Luruper Chaussee
149, 22761, Hamburg, Germany}
\emailAdd{cduhr@uni-bonn.de}
\emailAdd{pooja.mukherjee@desy.de}
\emailAdd{avasquez@uni-bonn.de}

\abstract{ 
We present a new \mathematica\ package that provides a platform to perform multi-loop computations. {\sc Anatar} integrates several existing tools designed for higher-order computations. In particular, it uses \qgraf\ to generate Feynman diagrams and relies on \form\ to perform the Dirac and colour algebras. This makes it very efficient without compromising user-friendliness and flexibility. In addition, \anatar\ is equipped with functionalities to manipulate the generated amplitudes and to extract scalar form factors. The resulting expressions can be mapped to integral families that can subsequently be reduced to master integrals via dedicated interfaces to codes for integration-by-parts reduction. Support for large classes of BSM models with the same gauge group as the SM is provided, and in particular {\sc Anatar} is optimised for computations in effective field theories. We review and introduce the main functionalities of {\sc Anatar} and illustrate its use for phenomenological applications.
 
}

\preprint{DESY-25-117, BONN-TH-2025-28 }

\begin{document}
\BgThispage
\maketitle

\section{Introduction} 
\label{sec:intro}
% !TEX root = ANATAR.tex

Scattering amplitudes are the backbone of almost all computations in Quantum Field Theory, including the computation of phenomenologically relevant predictions for collider and gravitational wave observables. It is typically not possible to evaluate scattering amplitudes exactly, but one resorts instead to perturbation theory, where the amplitudes are expanded in the small coupling constants of the theory. For example, if the theory contains a single coupling constant $\alpha$, then we can expand the amplitude $\cA$ perturbatively as
\beq
\cA = \alpha^m\,\sum_{L=0}^\infty\alpha^L\,\cA^{(L)}\,,
\eeq
where $\alpha^m$ is the coupling dependence of the leading-order amplitude. The amplitude depends on all the quantum numbers of the $N$ external particles that take part in the scattering process. In particular, it is a function of the four-momenta $p_i$ of the scattering particles. In the following we assume that all particles are in-going, and momentum conservation takes the form $\sum_{i=1}^Np_i=0$. 
 
 The standard textbook approach to compute perturbative  amplitudes are Feynman diagrams, where the $L^{\textrm{th}}$ perturbative coefficient is expressed as a sum over all $L$-loop Feynman diagrams with the prescribed external legs,
 \beq
 \cA^{(L)} = \sum_i\cD_i\,,
 \eeq
 where $\cD_i$ denotes the analytic expression for the $i^{\textrm{th}}$ Feynman diagram. Beyond tree-level, Feynman diagrams involve the integration over the $L$ loop momenta not fixed by momentum conservation. These so-called \emph{Feynman integrals} typically diverge and need to be regulated. The commonly-used regularisation scheme is dimensional regularisation~\cite{THOOFT1972189,Cicuta:1972jf,Bollini:1972ui}, where the computation is performed in $D=4-2\eps$ dimensions, and the divergences arise as poles in the dimensional regulator $\eps$. 
 
 In the following we assume that all external particles are scalars, vectors or Dirac fermions. We denote the numbers of external scalar, vector and fermion pairs by $S$, $V$ and $F$ respectively ($N=S+V+2F$), and we have,
 \beq
 \cD_i =  \widehat{\cD}_{i,{\bf s r}}^{\boldsymbol\mu} \prod_{v=1}^V\varepsilon_{\mu_v}\,\prod_{f=1}^F\overline{v}_{s_f}u_{r_f}\,,
\eeq
where $\varepsilon_{\mu}$, $\overline{v}_{s}$ and $u_{r}$ are the external polarisation vectors and Dirac spinors, and ${\boldsymbol{\mu}}=(\mu_1,\ldots, \mu_V)$, ${\bf s}= (s_1,\ldots,s_F)$ and ${\bf r}= (r_1,\ldots,r_F)$, and a sum over repeated indices is understood. The quantity $\widehat{\cD}_{i,{\bf s r}}^{\boldsymbol\mu}$ represents the Feynman diagram $ \cD_i$ where the external polarisation vectors and spinors have been amputated. Correspondingly, it is a tensor with $V$ Lorentz indices and $2F$ spin indices. 
If the external particles carry some gauge charges, then the Feynman diagrams are also tensors under the gauge group. For example, in applications to collider physics, one often encounters the situation where  the external particles transform in the fundamental or adjoint representations of the gauge group SU$(N_c)$ describing the strong interactions (with $N_c=3$). In the following, we collectively denote the external gauge indices by ${\bf a}$.
We can then define the scattering amplitude with external polarisation vectors and spinors removed as
 \beq
 \widehat{\cA}^{(L)\boldsymbol{\mu},{\bf a}}_{\bf sr} = \sum_i\widehat{\cD}_{i,{\bf sr}}^{\boldsymbol{\mu},{\bf a}}\,.
 \eeq
Each amputated diagram $\widehat{\cD}_{i,{\bf s r}}^{\boldsymbol\mu,{\bf a}}$ can be expressed as an $L$-loop tensor integral. 

It is usually preferable to work with scalar quantities rather than tensors. There are various algorithms to reduce tensor integrals to scalars, cf.,~e.g., refs.~\cite{Passarino:1978jh,Tarasov:1996bz,babis_thesis}. An alternative approach, typically employed for $L\ge 2$, consists of constructing a set of tensors $T_{i,{\bf s r}}^{\boldsymbol\mu,{\bf a}}$ and their dual projectors $P^{{\bf s r}}_{j,\boldsymbol\mu,{\bf a}}$ such that
\beq
T_{i,{\bf s r}}^{\boldsymbol\mu,{\bf a}} \,P^{{\bf s r}}_{j,\boldsymbol\mu,{\bf a}} = \delta_{ij}\,,
\eeq
where again a sum over repeated indices is understood.
The tensors and projectors only depend on the external particles, and in particular they are independent of the number of loops $L$.
There are various approaches to construct tensors for processes involving external vectors and fermions, cf.,~e.g.,~refs.~\cite{Peraro:2020sfm,Anastasiou:2023koq,Goode:2024mci}. Note that, when working in dimensional regularisation, it is important to construct tensors and projectors in $D$ dimensions. The precise form of the tensors and the projectors may then differ depending on the variant of dimensional regularisation used. For example, the tensors and projectors may be different in conventional dimensional regularisation (CDR), where all quantities are considered $D$-dimensional, or the 't Hooft-Veltman (tHV) scheme, where external particles are constrained to lie in four dimensions. If the amplitude is a tensor transforming in some representation of some gauge group, we can also contruct tensors and projectors for those indices.

Once a suitable set of tensors and projectors has been identified, it is possible to write the amplitude as a linear combination,
 \beq\label{eq:workflow_decomp}
\cA^{(L)} = \sum_j\mathcal{T}_j\,{\cF}_j^{(L)}\,,
 \eeq
 where the $L$-loop scalar form factors ${\cF}_j^{(L)}$ are obtained by acting with the projectors on the amputated amplitudes,
 \beq\label{eq:FF_def}
 {\cF}_j^{(L)} = P_{j,\boldsymbol{\mu},{\bf a}}^{\bf sr}\,\widehat{\cA}_{\bf sr}^{(L)\boldsymbol{\mu},{\bf a}}\,.
 \eeq
The dependence on the external polarisation vectors and spinors is captured via the contraction with the tensors, which are rational functions that only depend on the external legs,
\beq
\mathcal{T}_j =  T_{j,{\bf s r}}^{\boldsymbol\mu,{\bf a}} \prod_{v=1}^V\varepsilon_{\mu_v}\,\prod_{f=1}^F\overline{v}_{s_f}u_{r_f}\,.
\eeq

It is often not required to evaluate the form factors for the complete set of tensors, and only a subset of them contribute to a given physical observables. An example of this are scattering amplitudes with external on-shell gauge bosons. It is well known that massless gauge bosons in $D$ dimensions only have $(D-2)$ physical transverse degrees of freedom. Any tensor that vanishes when contracted with a polarisation vector corresponding to a physical polarisation state will not contribute to the scattering amplitude.

The previous step allows one to reduce the problem of computing a scattering amplitude to the computation of $L$-loop scalar form factors $\cF_j^{(L)}$. Schematically, each form factor has the form of an $L$-loop scalar Feynman integral,
\beq
\cF_j^{(L)} = \int \left(\prod_{l=1}^{L}\frac{\rd^Dk_l}{(2\pi)^D}\right)\,\frac{\cN_j(k_l\cdot k_m,k_l\cdot p_m)}{D_1\cdots D_r}\,,
\eeq
where $\cN_j(k_l\cdot k_m,k_l\cdot p_m)$ is a polynomial in scalar products involving loop and/or external momenta and the $D_k$ are (inverse) propagators,\footnote{Depending on the application, one may also encounter linear propagators.}
\beq
D_k = q_k^2-m_k^2+i\varepsilon\,,
\eeq
with $q_k$ a linear combination of the loop momenta $k_l$ and the external momenta $p_i$. The computation of scalar Feynman integrals is typically a challenging task, and a very active area of research (cf.,~e.g.,~refs.~\cite{Caola:2022ayt,Bourjaily:2022bwx} for recent reviews). It is useful to reduce the number of Feynman integrals that need to be evaluated to a minimum. As a first step, the integrals are sorted into different \emph{integral families}, often called \emph{topologies}, that only differ by the exponents of the propagators (see,~e.g.,~ref.~\cite{Pak:2011xt}). A member of a given integral family $F_i$ then takes the form
\beq
I^{F_i}_{\boldsymbol{\nu}}= \int \left(\prod_{l=1}^{L}\frac{\rd^Dk_l}{(2\pi)^D}\right)\,\frac{1}{D_1^{\nu_1}\cdots D_p^{\nu_p}}\,,\qquad \boldsymbol{\nu} = (\nu_1,\ldots,\nu_p) \in \mathbb{Z}^p\,, \label{eq:ScalarFI}
\eeq
where the denominators $D_k$ now include the inverse propagators, but also \emph{irreducible scalar products (ISPs)} that cannot be written as linear combinations of inverse propagators. The members of a given integral family generate a vector space, and it is well known that every integral family admits a finite basis, called \emph{master integrals}~\cite{Smirnov:2010hn, Bitoun:2017nre}. The relations between the different members of a family can be obtained from the so-called \emph{integration-by-parts} (IBP) identities~\cite{Tkachov:1981wb,Chetyrkin:1981qh}. Various sophisticated and algorithmic approaches to solve the system of IBP relations have been proposed~\cite{Laporta:2000dsw,Smirnov:2005ky,Smirnov:2006tz,Lee:2008tj,Schabinger:2011dz,vonManteuffel:2014ixa,Georgoudis:2016wff,Peraro:2016wsq,Peraro:2019svx,Barakat:2022qlc,Wu:2023upw,Mokrov:2023vva,Belitsky:2024jhe,Smirnov:2024onl,Chestnov:2024mnw,Wu:2025aeg}, and there are several public computer codes available for reducing Feynman integrals to a basis of master integrals~\cite{Anastasiou:2004vj,Smirnov:2008iw,Studerus:2009ye,vonManteuffel:2012np,Lee:2012cn,Lee:2013mka,Maierhofer:2017gsa,Klappert:2020nbg,Smirnov:2023yhb,Wu:2025aeg}. 

Once a reduction of all integrals to a set of master integrals has been achieved, the problem of evaluating the scalar form factors is reduced to the problem of computing the master integrals, either analytically or numerically. This is often a very complicated task in itself. Currently, there is no general strategy how to evaluate a given set of multi-loop Feynman integrals. However, it is understood how to systematically address all other steps and how to decompose a given scattering amplitude into scalar form factors expressed as linear combinations of master integrals.\footnote{Though in practice, some of the currently available algorithms  may require a prohibitively large run time or memory, depending on the complexity of the integrals involved.} We have already mentioned that there are various public codes to reduce a given integral family to a set of master integrals in an automated way. In addition, there are codes that allow the user to generate the multi-loop Feynman diagrams for a given process~\cite{qgraf,Hahn:2000kx}. The resulting analytic expressions can then be manipulated and processed using dedicated computer algebra packages~\cite{Vermaseren:2000nd,Hahn:2004rf,Hahn:2006zy,Kuipers:2012rf,Ruijl:2017dtg,Shtabovenko:2016sxi,Shtabovenko:2020gxv,Shtabovenko:2023idz,Hahn:2016ebn,symbolica,Mistlberger:2025ksa}.

Despite the high level of automation for the individual steps in the computation of multi-loop Feynman integrals, typically the individual codes cannot be seamlessly used within a common framework, requiring careful interfacing of different tools. Only very few efforts have been made to create a platform where all the different ingredients for multi-loop computations, from the diagram generation to the reduction to master integrals, are easily accessible to the user, cf.,~e.g.,~refs.~\cite{Fontes:2019wqh,Fontes:2021iue,Fontes:2025svw,Hahn:2000kx,Hahn:1998yk,Chen:2024xwt}. Moreover, to generate diagrams for new models and/or involving higher-dimensional operators, the user needs to implement the interaction vertices and the fields manually, which is a time-consuming and error-prone process. 

The purpose of this paper is to introduce \ant, a \mathematica\ package that provides a platform to perform multi-loop computations. \anatar\ can be used to perform most of the steps previously described in a single \mathematica\ environment. It contains a set of dedicated interfaces to common other packages for multi-loop computations, and thus provides a platform where these tools can be seamlessly combined into a single framework. For example, \anatar\ is interfaced to \qgraf~\cite{qgraf}, and the user may run \qgraf\ from within \anatar. The Feynman diagrams created by \qgraf\ are then transformed into analytic expressions, which are simplified using a dedicated interface to \form~\cite{Vermaseren:2000nd,Ruijl:2017dtg}. The model files required to run \qgraf\ can be generated directly from \feynrules~\cite{Christensen:2008py,Alloul:2013bka} via a newly developed interface that translates the vertices obtained by \feynrules\ into input files for \qgraf\ and \anatar. Once the analytic expressions for the individual Feynman diagrams are available, the user may decide to process them further in \mathematica. In particular, the user may define projectors and apply them to the diagrams to obtain scalar form factors. Finally, \anatar\ can sort the scalar form factors into scalar integral families, which can then be reduced to master integrals using available codes for IBP reduction. \anatar\ contains interfaces to \kira~\cite{Maierhofer:2017gsa,Klappert:2020nbg} and \litered~\cite{Lee:2012cn,Lee:2013mka} and allows the user to call these codes directly from within \anatar, or to just write out the input files needed to run these codes.

The paper is organised as follows: In section~\ref{sec:anatar} we discuss the main features of \anatar, in particular how to install the package and how to use it to generate and manipulate analytic expressions for Feynman diagrams. In section~\ref{sec:form_amps} we discuss how we can run \anatar\ to obtain scalar form factors and to reduce them to master integrals using \kira\ or \litered. In section~\ref{sec:validation} we discuss the list of known processes that we have recomputed to validate our code, and in section~\ref{sec:example} we illustrate the usage of the code on a non-trivial example. Finally, in section~\ref{sec:conclusion} we draw our conclusions. We provide three appendices: one detailing the structure of the input model files for \anatar, another explaining how to generate these files using \feynrules, and a third describing how \anatar\ handles the $\gamma_5$ matrix in dimensional regularisation.

\section{The \anatar\ package}
\label{sec:anatar}
In this section we describe the main usage and the functionalities of the \anatar\ package. 
We focus on the generation and manipulation of amplitudes in the Standard Model (SM) of particle physics. Features of the package that can be obtained via interfacing to codes 
for IBP reduction via \litered~\cite{Lee:2012cn,Lee:2013mka} or  \kira~\cite{Klappert:2020nbg,Maierhofer:2017gsa}  will be described in subsequent sections.
% !TEX root = anatar.tex
\subsection{Installation}
\label{sec:installation}

\anatar\ is available from the git repository at 
\begin{center}
{\tt https:}$\,${\tt //github.com/ANATAR-hep/ANATAR}
\end{center}
It is possible to download a compressed file with the latest version of the repository from this gitlab URL. Alternatively, \anatar\ can be obtained by cloning the repository using git, e.g.,~by issuing the following command in any shell that has the \texttt{git} command available:
\begin{verbatim}
    git clone https://github.com/ANATAR-hep/ANATAR.git
\end{verbatim}
This will clone the \anatar\ repository into the current working directory.

Afterwards \anatar\  can be loaded into an active \mathematica\ session by issuing:
\exbox{
\inline{1}{& {\tt \$AnatarPath=SetDirectory[<path>];}}\\
\inline{2}{& {\tt <<Anatar`;}}
}
where {\tt <path>} is the path to the folder that contains the file {\tt Anatar.m}. 

At this point we need to make an important comment. As mentioned above, \anatar\ relies on \qgraf\ to generate the Feynman graphs of a process and on \form\ to perform algebraic manipulations. Moreover, it is possible to call \kira\ or \litered\ (which itself uses \fermatica~\cite{fermatica}) to reduce families of Feynman integrals to master integrals. Therefore, it is mandatory that \anatar\ knows how to call these external programs. In the following, we explain how to ensure that \qgraf\ and \form\ can be called correctly from within \anatar. We will discuss \kira\ and \litered\ in section~\ref{sec:IBP}, where the interfaces to those programs are introduced.

For \anatar~to be able to generate Feynman diagrams with \qgraf, a working copy of \qgraf~must be installed on the machine. The path to the folder that contains the executable {\tt qgraf} must be given by the variable {\tt \$QGrafPath}. Furthermore, a new style file for \qgraf, namely {\tt array\_FR.sty}, is required to ensure that the diagrams can be read by \ant. This file is provided in the cloned folder. Thus, on Unix, Linux, or MacOS systems it is enough to do
\begin{verbatim}
    mv array_FR.sty <QGrafPath>
\end{verbatim}
with {\tt <QGrafPath>} being the path where the {\tt qgraf} exectutable is found. 

The analytic expressions for the Feynman diagrams are manipulated with \form\ \cite{Ruijl:2017dtg}. For this \form\ needs to be installed on the machine and must be callable from the shell by issuing the {\tt form} command. When loading \anatar\ into the \mathematica\ kernel, the package checks if the \form~executable can be called. If not, an error message is issued.

\subsection{Amplitude generation}
\label{sec:amplitudes}
% !TEX root = anatar.tex

We now describe the main usage of \anatar, that is, how \anatar~can be used to generate and manipulate analytic expressions for scattering amplitudes.

The first step is to load the model files containing the definitions (e.g., fields and interaction vertices) for the model under consideration. The model files are stored inside subfolders in the folder {\tt "Models"} of the \anatar\ directory. 
The concrete syntax of the model files is irrelevant for understanding the generation of the amplitudes. Here, it suffices to say that \anatar\ is shipped with a set of complete models\footnote{We also provide a \feynrules~\cite{Alloul:2013bka} interface which can assist in the creation of new models, see Appendix \ref{sec:modelGen}.}, listed in Table \ref{tab:models}.
For example, the model {\tt QED} can be loaded by issuing the command
\exbox{
\inline{1}{& {\tt LoadModel["QED"];}}
}
\begin{table}
\bgfb
\multicolumn{2}{c}{\textbf{Table~\ref{tab:models}: Available models in \ant. }}\\
\\
\tt{{\tt sm}} & The SM implementation of \anatar~in both Feynman and unitary gauge. The QCD sector offers the switch between Feynman and covariant gauge. The model treats the bottom and top quarks as massive, while all other quarks are considered massless.  The chosen parameters for the option {\tt CouplingsOrder} are {\tt gs, ee} and {\tt MT}.
\vspace{0.2cm}
\\
{\tt QED} & The QED implementation of \anatar. It includes the electron, muon, tau and top quark fields and their interactions mediated by the photon. The chosen parameter for the option {\tt CouplingsOrder} is {\tt ee}.
\vspace{0.2cm}
\\
{\tt chromo} & An \ant~implementation of the SM with some dimension-six operators added, in both Feynman and unitary gauges. Only fields involved in those effective operators are kept. The model treats the bottom and top quarks as massive, while all other quarks are considered massless. The QCD sector offers the switch between Feynman and covariant gauge. The model includes the operators $\mathcal{O}_{tG}$, $\mathcal{O}_{t\phi}$ and $\mathcal{O}_{\phi G}$ as defined in ref.~\cite{Degrande_2021}. The chosen parameters for the option {\tt CouplingsOrder} are {\tt gs, MT} and {\tt cEFT}, with the last one standing as a flag for any effective coupling and which should be set to one after amplitude generation.
\egfb
\phantomcaption{\label{tab:models}}
\end{table}
Information about a model (after it has been loaded) can be obtained at run time by issuing the command
\newline\newline
{\tt AnatarModel[]}
\newline\newline
This creates a \mathematica\ association that summarises the main properties of the model, including the particles defined in the model. More information about a specific field ${\tt psi}$ defined in the model can be obtained by issuing the command
\newline\newline
{\tt Particle[psi]}
\newline\newline
For more details on the structure of the model files, we refer to appendix~\ref{sec:models}, where we also describe a \feynrules\ interface to automatically generate new \anatar\ models directly from a Lagrangian. 

After loading a model, scattering amplitudes can be generated by issuing the following command in an active \mathematica\ session:
\newline
\newline
{\tt GenerateAmplitude[\{psi1,psi2,...\} -> \{psiA,psiB,...\},} \emph{options}{\tt]},
\newline
\newline
where \emph{options} refers to a set of optional arguments described in Table~\ref{tab:OptionsAmplitude}. In particular, the number of loops of the generated graphs is provided by the option {\tt Loops} (the default value being 0).
Here, the incoming and outgoing fields are collected in the lists ${\tt \{psi1,psi2,...\}}$ and ${\tt \{psiA,psiB,...\}}$. \anatar\ then runs \qgraf\ behind the scenes to generate all relevant Feynman diagrams. The output files produced by \qgraf\ are saved in a folder dedicated to the process, and subsequently read into \mathematica. The output of \qgraf~is purely topological and does not yet provide analytic expressions for the Feynman diagrams. \anatar\ converts the topological information generated by \qgraf\ into analytic expressions in \mathematica. The output is an association which summarises the main properties of the amplitude, and also contains the analytic expressions for the individual Feynman diagrams. The output is also written to file. The name of the output folder is provided by the option {\tt OutputName}. If omitted, an automatic name is created according to the lists of in- and outgoing fields.
Analytic expressions for scattering amplitudes can be quite large.
To save disk space, we use the Unix/Linux {\tt gunzip} command to compress the output produced by \form~during the generation of scattering amplitudes. Since this command is built into \form, any intermediate files produced during its execution are automatically saved in compressed format. On some systems {\tt gunzip} may not be available or functional. In such cases, a warning message will appear during the initialisation of the package. This warning is harmless and does not affect the execution or outcome of the subsequent computations.

We stress that \anatar\ automatically assigns symbols for the loop and external momenta. The momenta of the external particles are called {\tt p1}, {\tt p2}, $\ldots$, in the order in which they appear in the joined list ${\tt \{psi1, psi2,..., psiA, psiB,...\}}$. In addition to these external momenta, an $L$-loop amplitude depends on $L$ momenta {\tt k}1, $\ldots$, {\tt k}$L$. The user should avoid using these variable names for other quantities.

As an example, let us consider top-pair production in electron-positron annihilation, $e^+ e^- \rightarrow t\bar{t}$, in QED at tree level. The electron is kept massless. This process can be generated via the command:
{\small
\exbox{
\inline{2}{& {\tt EETT0 =  GenerateAmplitude[\{e,ebar\}->\{t,tbar\}, Dimension->4]}}\\
\breakline\\
\outline{2}{& {\tt <|Process->\{\{e[p1],ebar[p2]\} -> \{t[p3],tbar[p4]\}\}, } \\
 & {\tt \; Total->\{1;;1\}, Name->"eebar\_ttbar", LoopOrder->0,} \\
 & {\tt \; Kinematics->\{p1\pw2->0,p1.p1->0,p2\pw2->0,p2.p2->0,p3\pw2->MT\pw2,}\\
 & {\tt \; \;\;p3.p3->MT\pw2,(p3.p3)\pw(-1)->MT\pw(-2),p4\pw2->MT\pw2,p4.p4->MT\pw2,}\\
 & {\tt \; \;\;(p4.p4)\pw(-1)->MT\pw(-2),p1.p2->S/2,p1.p3->(MT\pw2-T)/2,}\\
 & {\tt \; \;\;p2.p3->(MT\pw2-U)/2,(p1.p2)\pw(-1)->2/S,(p1.p3)\pw(-1)->}\\
 & {\tt \; \;\;2/(MT\pw2-T),(p2.p3)\pw(-1)->2/(MT\pw2-U),p4->p1+p2-p3\},}\\
 & {\tt \; Amplitudes->\{DiagramID[1] -> (-I)*GC2*GC3*Den[-p1-p2,0]*}  \\
 & {\tt \; \;\;GammaM[iLor2,Spin2,Spin4]*GammaM[iLor2,Spin3,Spin1]*}\\
 & {\tt \; \;\;SpinorU[-1,p1,0,Spin1]*SpinorUbar[-2,p3,MT,Spin2,Colour4]*} \\
 & {\tt \; \;\;SpinorV[-4,p4,MT,Spin4,Colour4]*SpinorVbar[-3,p2,0,Spin3]\}|>}  
 }
}
}
The analytic expression for the amplitude is given by the {\tt Amplitudes} key in the association, and can be accessed in the standard way as {\tt EETT0[Amplitudes]}. Its value is a list of rules whose entries correspond to the analytic expressions for the individual Feynman diagrams generated by \qgraf. The result for the $i^{\textrm{th}}$ diagram is tagged as {\tt DiagramID[}\emph{i}{\tt]}. In the example above, the only tag is {\tt DiagramID[1]}.

In general, the expression for the amplitude is given in terms of the objects listed in Table \ref{tab:objectsAmp}. We notice that the spinor indices are kept explicit, so that the order of the printed objects in the output does not matter. 
Although not relevant for the example above, in \ant~the $\gamma_5$-scheme in $D$-dimensions is Larin's prescription \cite{Akyeampong:1973xi,Larin:1991tj}. For more details, we refer to appendix \ref{sec:gamma5}.
%%%%%%%%%%%%%%%%%%%%%%%%%%%%%%%%%%%%%%%%%%%%%%%%%%%%%%%%%%%%%%%%%%%%%%%
\begin{table}
\bgfb
\multicolumn{2}{c}{\textbf{Table~\ref{tab:objectsAmp}: Syntax used in \ant. }}\\
\\
{\tt  d } & Space-time dimension $D$.
\\
{\tt  p1,p2,...} & External momenta $p_1$, $p_2,$,...
\\
{\tt  k1,k2,...} & Loop momenta $k_1$, $k_2,$,...
\\
{\tt  p1[Lor1],...} & Four-vector $p_1^{\rm Lor1}$. 
\\
{\tt  p1.k1,p2.k1,...} & Scalar products of four-vectors, $p_1\cdot k_1$, $p_2\cdot k_1$,...
\\
{\tt  Den[p1,m] } & The scalar propagator $\frac{1}{p_1^2-m^2}$.
\\
{\tt  GammaM[1,Spin1,Spin2] } & Identity element in the spinor space $ \unit_{\rm  Spin1,Spin2}$. 
\\
{\tt  GammaM[Lor1,Spin1,Spin2] } & The Dirac matrix $ \gamma^{\rm Lor1}_{\rm  Spin1,Spin2}$. 
\\
{\tt  GammaM[p, Spin1,Spin2] } & Contractions of Dirac matrices and momenta, $ \slashed p_{\rm  Spin1,Spin2}$. 
\\
{\tt  GammaM[5,Spin1,Spin2] } & The matrix $\gamma^{5}_{\rm  Spin1,Spin2}$. 
\\
{\tt SpinorU[-1,$p_1$,m,Spin1,Color1] } & The Dirac spinor $u_{\rm Spin1}^{\tt Color1}(p_1,m)$, associated with the fermion of mass $m$ and colour {\tt Colour1}. Similar conventions follow {\tt SpinorUbar}, {\tt SpinorV} and {\tt SpinorVbar}. 
\\
{\tt SpinorU[-1,$p_1$,m,Spin1] } & The Dirac spinor $u_{\rm Spin1}(p_1,m)$, associated with the fermion of mass $m$. Similar conventions follow {\tt SpinorUbar}, {\tt SpinorV} and {\tt SpinorVbar}. For color singlet fields the {\tt Color1} index is omitted.
\\
{\tt PolV[Lor1,$p_1$,0] } & Polarization vector $\varepsilon_{\rm Lor1}(p_1,0)$, associated with the external massless vector bosons, cf. eq.~\eqref{eq:polsumMassless}.
\\
{\tt PolV[Lor1,$p_1$,m] } & Polarization vector $\varepsilon_{\rm Lor1}(p_1,m)$, associated with the external vector boson of mass $m$, cf. eq.~\eqref{eq:polsumMassive}.
\\
{\tt PolV[Lor1,$p_1$,0,Gluon1] } & Polarization vector $\varepsilon^{\tt Gluon1}_{\rm Lor1}(p_1,0)$, associated with the external massless vector boson of colour index {\tt Gluon1}, cf. eq.~\eqref{eq:polsumMassless2}.
\\
{\tt f[Gluon1, Gluon2, Gluon3] } & The QCD structure constant $ f^{\tt Gluon1\,Gluon2\,Gluon3} $.
\\
{\tt T[Colour1,Colour2,Gluon1]} &  The QCD generator $T^{\tt Gluon1}_{\tt Colour1 \, Colour2} $.
\\
{\tt LeviCivita[Lor1,Lor2,Lor3,Lor4]} &  Levi-Civita symbol $\varepsilon_{\tt Lor1,Lor2,Lor3,Lor4}$ with the convention used $\varepsilon_{0123}=-\varepsilon^{0123}=+1$. 
\\
{\tt Metric[i,j]} &  Symbol for the Kronecker's delta and Minkowski metric,   $g_{\tt i,j}$ and $\delta_{\tt i,j}$. 
\\
{\tt nn1,nn2,...} &  External four-vectors introduced through polarization sums, cf. eq.~\eqref{eq:polsumMassless2}.
\\
{\tt TF,CA,CF,Nc} &  Colour factors of the SU($N_c$) algebra. 
\egfb
\phantomcaption{\label{tab:objectsAmp}}
\end{table}

%%%%%%%%%%%%%%%%%%%%%%%%%%%%%%%%%%%%%%%%%%%%%%%%%%%%%%%%%%%%%%%%%%%%%%%

Let us now describe some basic operations that we can perform on the generated amplitudes. First, we can ask \anatar\ to sum the expressions for the individual diagrams by issuing
\exbox{
\inline{3}{&{\tt EETT0Sum = SumDiagrams[EETT0];}}
}
The result is a new association identical to the old one, but where the value of the {\tt Amplitudes} key now contains a single entry, namely
\newline\newline
{\tt \{DiagramID[Sum] -> ...\}}
\newline\newline
where `{\tt ...}' represents the sum of the analytic expressions for the individual diagrams. More generally,  the command
\exbox{
\inline{4}{& {\tt EETT0FuncMap = AmplitudesMap[EETT0, func];}}
}
returns an association identical to {\tt EETT0}, but with the value of the {\tt Amplitudes} key replaced by
\newline\newline
{\tt \{DiagramID[1] -> func[...], ...\}}
\newline\newline
In this way, it is possible to apply standard algebraic manipulations directly to the associations.
We can also act with the function {\tt AmplitudesMap[]} after having summed the diagrams. The functions {\tt AmplitudesSimplify[]} and {\tt AmplitudesFullSimplify[]} are equivalent to calling {\tt AmplitudesMap[]} with the second argument {\tt func} equal to {\tt Simplify} or {\tt FullSimplify}.

For complicated models, the analytic expressions of the couplings constants appearing in the vertices can be complicated functions of the parameters of the model. Having the correct algebraic expressions for the couplings is important, e.g., to ensure gauge invariance of the amplitudes. At the same time, these analytic expressions may render the expressions for the vertices lengthy and unwieldy. For this reason, it is often useful to introduce abbreviations for the coupling constants in the vertices.\footnote{If the model was generated with \feynrules, these abbreviations are introduced automatically, see appendix~\ref{sec:models}.} It is possible to insert the expressions for the abbreviations in the value of the {\tt Amplitudes} key using the following command:
\exbox{
\inline{5}{& {\tt EETT0Abbr = InsertCouplings[EETT0];}}
}

%%%%%%%%%%%%%%%%%%%%%%%%%%%%%%%%%%
\subsection{Options available for the amplitude generation}
%%%%%%%%%%%%%%%%%%%%%%%%%%%%%%%%%%

The example above is a tree-level process in QED, and thus it is relatively simple. \ant~is designed to deal with more involved scattering amplitudes. For this reason, the generation of amplitudes comes equipped with a broad range of options. The complete set of  options of {\tt GenerateAmplitudes[]} are listed in Table \ref{tab:OptionsAmplitude}. The first set of options generally apply to the generation of Feynman diagrams. The second set corresponds to options that allows a tailored selection of diagrams and amplitudes.
%%%%%%%%%%%%%%%%%%%%%%%%%%%%%%%%%%%%%%%%%%%%%%%%%%%%%%%%%%%%%%%%%%%%%%%
\begin{table}
\bgfb
\multicolumn{2}{c}{\textbf{Table~\ref{tab:OptionsAmplitude}: Options of {\tt GenerateAmplitude} }}\\
\\
{\tt Model} & Loads the selected model.
\\
{\tt OutputName} & Choose the name of the amplitude and the output folder. The default name is set according to the external fields.
\\
{\tt Loops} & Defines the number of loops in the amplitude. Default value is {\tt 0}.
\\
{\tt Dimension} & Selects the dimension for a process. The default value is  {\tt d} for both tree-level and higher-order amplitudes.
\\
{\tt PolarizationTrim} & When set to {\tt True}, allows the removal of polarization vectors and spinors. Default value is {\tt False}. 
\\
{\tt Gauges} & Specifies the gauge of preference, the default choice is {\tt \{"QCD"->"Feynman","EW"->"Feynman"\}}.
\\
{\tt SubstitutionRules} & Replacement rules can be given to the generation of the amplitude. These expressions may include tensor structures such as Dirac matrices, spinors, colour matrices or numerical factors.
\\
{\tt SimplifiedOutput} & When set to {\tt True} performs simplifications on the colour and Lorentz structures. Default value is {\tt False}.
\\
{\tt RecycleAmplitude} & When set to {\tt True} imports the {\tt Amp\_l.m} file found in the output folder name provided, this without generating the amplitudes again. The default value is {\tt False}. 
\\
{\tt SelectAmplitudes} & Allows the selection of the diagrams to be processed under Feynman Rules. Default value is {\tt All}.
\\
{\tt CouplingsOrder} & Specifies the coupling orders for the diagrams where a generic coupling is selected to appear from powers minimum to maximum. Default value is {\tt None}.
\\
{\tt SelectCouplingsOrder} &  Selects terms in the amplitudes, yielding the coefficient of a given parameter to a specified power. Default value is {\tt All}.
\\
{\tt QGrafOptions} &  Allows the selection of diagrams according to the specified graph topologies. Default value is {\tt None}.
\\
{\tt RemovePropagator} & Forbids the insertion of propagators associated to the specified fields. Default value is {\tt None}.
\\
{\tt QGPartition} & Restricts the diagram generation according to the partitions in \qgraf. Default value is {\tt None}.
\\
{\tt QGMessage} & Assigns the name of the log-file saving warning and error messages coming from running \qgraf. Default value is {\tt None}.
\\
{\tt OffShell} & Sets the external momenta off-shell. Default setting is {\tt False}.
\\
{\tt OnlyDiagrams} & When set to {\tt True} generates only the diagrams without generating the respective amplitudes. Default setting is {\tt False}.
\egfb
\phantomcaption{\label{tab:OptionsAmplitude}}
\end{table}
%%%%%%%%%%%%%%%%%%%%%%%%%%%%%%%%%%%%%%%%%%%%%%%%%%%%%%%%%%%%%%%%%%%%%%%

Let us start by describing the options from the first set, which generically apply to the generation of amplitudes. The number of loops in the amplitude can be selected by including the option {\tt Loops}, so that when set to {\tt 0} tree-level diagrams are produced.

The {\tt Dimension} option allows the user to specify the space-time dimension used in the calculation. It can be set either to any even integer or to a symbolic value, with the default being the symbolic variable {\tt d}. All calculations until the point where the IBP reduction has been completed should be performed with the symbolic {\tt d}, regardless of whether they are at tree-level or higher-loop orders. Only after the IBP reduction has been completed should {\tt d} be replaced by an explicit expression of the form $m - n \epsilon$, where $m$ is an even integer and $\epsilon$ is the dimensional regulator. This separation ensures that the reduction algorithms remain general and independent of the specific dimensional expansion used later.

The option {\tt Model} allows the user to select a model for the given amplitude. Note that adding this option is equivalent to calling the {\tt LoadModel[]} function as described in section~\ref{sec:amplitudes}. If the model was loaded before, there is no need to load it again via the option {\tt Model}. However, it may be useful to load a different model within the same \anatar\ session. For example, this makes flexible the computation of interferences between a SM amplitude and the corresponding amplitude within a beyond the SM model. 

In applications it may be useful to have expressions for scattering amplitudes with the external polarisation vectors and Dirac spinors amputated. This can be achieved by setting the option {\tt PolarizationTrim} to {\tt True} (the default value is {\tt False}). In cases where the amplitude is subsequently contracted with suitable projectors, this option must be set to {\tt True}.  
 
 The option {\tt Gauges} offers the option to select between different gauges. For the QCD sector, {\tt "QCD"}, there is the option to switch between Feynman and Covariant gauges, called {\tt "Feynman"} and {\tt "RXi"}, respectively. For models with an electroweak sector, the options for {\tt "EW"} are the Feynman and Unitary gauges, labelled {\tt "Feynman"} and {\tt "Unitary"}, respectively. 
 
 The option {\tt SubstitutionRules} offers the possibility to add replacement rules that are processed in the \form\ code for the generation of the scattering amplitude. This has the advantage of better performance than the corresponding evaluation in \mathematica\ after the generation of the amplitude. In some cases, it could even be more convenient to set certain parameters to zero as soon as the expressions of the Feynman rules are inserted.

The second set of options allows the user to select specific diagrams, scattering amplitudes, or terms. To choose specific diagrams by their label number in the \qgraf~output, the option {\tt SelectAmplitudes} is suitable. The syntax is based on the \mathematica~notation of the function {\tt Part} ({\tt [[...]]}), so that, for example, {\tt SelectAmplitudes->\{m;;n\}} gives the diagrams {\tt m} through {\tt n} from the \qgraf\ output. This option is designed for computations that require the evaluation of a large number of Feynman diagrams, or for estimates of times for very involved computations, in which case one can select only the first diagram generated by \qgraf. Notice that this function generates all the diagrams via \qgraf~but only imports and evaluates the selected diagrams. 

The diagrams generated by \qgraf~can be restricted using the option {\tt CouplingsOrder}, which follows the syntax
\newline\newline
{\tt CouplingsOrder -> \{\{alpha1, min1, max1\},\{alpha2, min2, max2\},...\} }
\newline\newline
In this way, the generic couplings {\tt alpha}$x$ are specified to appear with powers ranging from {\tt min}$x$ to {\tt max}$x$ in the amplitudes. In cases where {\tt min}$x$ and {\tt max}$x$ are equal to the same value, say {\tt r}$x$, it is enough to write {\tt \{alpha}$x${\tt , r}$x${\tt \}}. It is important to note that this is done at the diagram generation level, before inserting the symbolic expressions of the Feynman rules. This means that the information on the power associated with each coupling for a given vertex is provided in the \qgraf~model. The option {\tt SelectCouplingsOrder} assists in cases where the coefficient of a symbol {\tt s}$x$ with specific power {\tt a}$x$ is required. In this case, the syntax is:
\newline\newline
{\tt SelectCouplingsOrder -> \{\{s1,a1\},\{s2,a2\},...\} }
\newline\newline
The options {\tt CouplingsOrder} and {\tt SelectCouplingsOrder} complement each other, excelling in cases where the SM vertices receive new terms from new-physics extensions. For example, let us consider the model {\tt chromo}, which contains modifications to the interactions between the gluon and the top quark, gluon self-interactions and to the top-Yukawa. Let us say that we are interested in the insertions of the top-gluon interaction to the usual triangle diagrams of the Higgs production via gluon fusion. We observe that the associated Feynman rule has terms coming from the SM, tagged {\tt gs}, and from a given EFT, tagged {\tt cEFT}. In the \qgraf~model this vertex will show as 
\newline\newline
{\tt [ tbar, t, G; gs=1, cEFT=1, MT=0 ] } 
\newline\newline
with {\tt MT=0} indicating that this coupling does not receive contributions from the top-Yukawa coupling. With this definition of the vertex, for a process like the Higgs production via gluon fusion, the specification of the coupling order must be 
\newline\newline
{\tt CouplingsOrder -> \{\{gs,2\},\{cEFT,3\},\{MT,1\}\} }
\newline\newline
Notice that the top-Yukawa coupling also contains a contribution of the order {\tt cEFT}. This leads to scattering amplitudes that include up to triple insertions of effective operators, which in many cases one does not want to consider. By adding the option 
\newline\newline
{\tt SelectCouplingsOrder -> \{\{cEFT,1\}\} }
\newline\newline
we ensure that only linear terms in the parameter {\tt cEFT} are treated at the amplitude generation. This proves to be very efficient for higher-dimensional operators, where one is typically interested in only a fixed number of insertions of these operators into the diagrams. Alternatively to the procedure above, diagrams could be generated without specifying any values for the option {\tt CouplingsOrder}, and then directly use {\tt SelectCouplingsOrder} to obtain linear terms. However, in the example above, this involves handling 20 diagrams, of which many are of no interest, given the goal of only checking for modifications induced by top-gluon interactions. For higher-loop computations, this might be increasingly inefficient.

There are several options offered by \qgraf~for the selection of diagrams by topologies of the graphs. Such options can be included by issuing {\tt QGrafOptions}. The possibilities are the following:
\begin{center}
\centering
\begin{tabular}{c c c}
   {\tt bipart}  & {\tt cycli} & {\tt onepi}\\
   {\tt nodiloop}  & {\tt \qquad\qquad noparallel\qquad\qquad} & {\tt noselfloop} \\
   {\tt onevi}  & {\tt onshellx} & {\tt nosnail}\\
\end{tabular}
\end{center}
These options also have duals, such as {\tt nonbipart}. Details on each of these options can be found in the \qgraf~documentation contained in its installation folder~\cite{qgraf}. Furthermore, diagrams with propagators associated to specific fields can also be discriminated. Using the option 
\newline\newline
{\tt RemovePropagator -> \{\{psi1, min1, max1\},\{psi2, min2, max2\},...\} }
\newline\newline
keeps diagrams with $m$ ocurencies of a propagator associated to the field {\tt psi}$x$, with {\tt min}$x\le m\le$ {\tt max}$x$. Similarly to the option {\tt CouplingsOrder}, the syntax {\tt \{psi}$x${\tt , r}$x${\tt \}} stands for cases with {\tt min}$x$ and {\tt max}$x$ equal to the same value {\tt r}$x$. Hence, setting both {\tt min}$x$ and {\tt max}$x$ to zero forbids any diagram with a propagator of the given field {\tt psi}$x$. The short syntax 
\newline\newline
{\tt RemovePropagator -> \{psi1, psi2,...\} }
\newline\newline
removes every diagram that contains any number of insertions of the fields {\tt psi}$x$.  Notice that this option could be of use to efficiently deal with processes where the light quarks contribute in the same manner. In such cases, it often suffices to keep one flavour while the remaining ones could be removed using the {\tt RemovePropagator} option. The resulting amplitudes should then be multiplied by the number of light quarks accordingly. Finally, the generation of diagrams can also be restricted by using the option {\tt QGPartition}, which does so according to the multiplicity of the types of vertices of the selected model. If the model has 3- and 4-point vertices, we can select the number of times, {\tt n3} and {\tt n4}, each vertex can be allowed in the generated graphs. The syntax in general follows
\newline\newline
{\tt QGPartition -> " 3\pw n3 \tt 4\pw n4 5\pw n5 6\pw n6 ... " }
\newline\newline
For example, setting {\tt n3} to {\tt 0} forbids any 3-point vertex in the graphs, while setting {\tt n4} to {\tt 1} generates graphs with only one insertion of 4-point vertices. Inequalities can be given by adding the symbols {\tt +} and {\tt -}, so that for the example 
\newline\newline
{\tt QGPartition -> " 3\pw (2+) \tt 4\pw 1 5\pw (1-) " }
\newline\newline
only graphs with at least two 3-point vertices, one 4-point vertex, and at most one 5-point vertex are produced. The option {\tt QGMessage} takes a string as input that indicates the name of the log file that contains warning errors and messages generated by \qgraf. 
%This could be useful for debugging a model file. By default \ant~does not generate such a file.

Finally, we make the following comment. So far we have only discussed the generation of on-shell scattering amplitudes. It is also possible to use \anatar\ to generate off-shell Green's functions. This is done by setting the option
\newline\newline
{\tt OffShell -> True}
\newline\newline
% Note that this instruction is not passed as an option of {\tt GenerateAmplitudes[]}, but it variable on its own.
Hence, when set to {\tt True}, all the subsequent computations are performed off-shell. Additionally, this fixes the value of the symbol {\tt SetOffShell}, so that all external momenta are taken off-shell in the subsequent computations as long as its value is {\tt True}. Effectively, this option treats expressions in terms of scalar products {\tt pi$\cdot$pj} and does not substitute the corresponding values of {\tt pi$\cdot$pi}. Thus, the only simplification executed from kinematics is momentum conservation. The only exceptions are two-point amplitudes where {\tt p1$\cdot$p1} is substituted by {\tt S}. Finally, the {\tt SetOffShell} option should be used only at the amplitude level, and not when squaring amplitudes.

%%%%%%%%%%%%%%%%%%%%%%%%%%%%%%%%%%
\subsection{Generation of squared amplitudes}
%%%%%%%%%%%%%%%%%%%%%%%%%%%%%%%%%%

So far we have described the generation of scattering amplitudes or off-shell Green's functions. 
It is possible to use \anatar\ to generate expressions for the conjugate and/or squared amplitudes, as we now describe.

We assume from here on that the amplitude at a given loop order for the process {\tt \{psi1,psi2,...\} -> \{psiA,psiB,...\}} has been generated with {\tt GenerateAmplitude[]}, and the output is stored in the association {\tt amp}.
%sec:amplitudes
To compute a squared amplitude, or more generally an interference between two amplitudes with these external fields, \ant~requires first to generate the conjugated amplitude at order $L'$ via the command: 
\newline
\newline
{\tt GenerateAmplitudeConjugate[\{psi1,psi2,...\} -> \{psiA,psiB,...\},} \emph{options}{\tt]}
\newline
\newline
where \emph{options} corresponds to the same set of options of {\tt GenerateAmplitude}, presented in Table \ref{tab:OptionsAmplitude}. This produces an association (which we call {\tt ampC}) in terms of the hermitian conjugates of the objects listed in Table \ref{tab:objectsAmp}. 
These non-Hermitian objects have the suffix `{\tt C}', so that, for example, the conjugate of a polarization vector is named {\tt PolVC}.
We stress that {\tt amp} and {\tt ampC} must not necessarily be complex conjugates of each other! For example, we may be interested in computing the interference between the $L$ and $L'$ loop amplitudes. For this reason, we prefer to keep the generation of amplitudes and their conjugates separate.

After the  amplitude (conjugate) {\tt amp} and {\tt ampC} have been generated, we may compute their inteference via
\newline
\newline
{\tt GenerateAmplitudeSquare[amp,ampC,}\emph{options}{\tt]}\,.
\newline
\newline
The list of available \emph{options} refers to a set of optional arguments described in Table~\ref{tab:AmplitudeSquareOptions}. Note that {\tt amp} and {\tt ampC} are the full on-shell scattering amplitudes, including the polarisation vectors. In particular, the option {\tt PolarizationTrim} should be set to {\tt False} (which is its default value). Note that {\tt GenerateAmplitudeSquare[]} automatically sums all the contributions from individual Feynman diagrams.

\anatar\ then performs the sum over spins for Dirac fermions using the usual formula
\begin{equation}
\sum_{s=\pm1}u_s(p)\overline{u}_s(p) = \slashed{p}+m\,,\qquad \sum_{s=\pm1}v_s(p)\overline{v}_s(p) = \slashed{p}-m\,.
\end{equation}
Summations over the polarisations of vector bosons are performed using the identities
\begin{align}
    \sum_{\lambda=\pm1} \varepsilon_{\lambda,\mu}(k,\lambda)^*\, \varepsilon_\nu(k,\lambda) & = - g_{\mu\nu}\,, \label{eq:polsumMassless} \\
    \sum_{\lambda=-1}^{+1} \varepsilon_\mu(k,\lambda,M)^*\, \varepsilon_\nu(k,\lambda,M) & = - g_{\mu\nu} + \frac{k_\mu k_\nu}{M^2}\,, \label{eq:polsumMassive}\\
    \sum_{\lambda=\pm1} \varepsilon_{\mu}^a(k,\lambda,n)^*\, \varepsilon^b_\nu(k,\lambda,n) & = \delta^{ab} \left( - g_{\mu\nu} + \frac{n_\mu k_\nu+k_\mu n_\nu}{n\cdot k} \right)\,, \label{eq:polsumMassless2} 
\end{align}
for abelian massless,  massive and non-abelian massless vector bosons, respectively. We highlight that averaging factors for the initial states are not included for any field type. Additionally, notice that in general for massless vector bosons the polarization sum is gauge-dependent, as is manifest through the appearance of the external four-vector $n$, represented as {\tt nn} in the output. The polarization sum in eq.~\eqref{eq:polsumMassless2} corresponds to the axial gauge, and the vector $n$ satisfies the relations $n\cdot\varepsilon = 0$, $n^2 = 0$ and $n\cdot k \neq 0$. We notice that for vector bosons corresponding to an abelian gauge group it is enough to take the polarization sum as in eq.~\eqref{eq:polsumMassless}. When set to {\tt True}, the option {\tt nnVectorSimplification} instructs \anatar\ to use momentum conservation to simplify squared amplitudes with external gluons to reduce them to a form that explicitly displays their gauge independence. However, this function, in its current form, may not be particularly efficient beyond the tree level. 
%%%%%%%%%%%%%%%%%%%%%%%%%%%%%%%%%%
\begin{table}
\bgfb
\multicolumn{2}{c}{\textbf{Table~\ref{tab:AmplitudeSquareOptions}: Options of {\tt GenerateAmplitudeSquare}}}\\
\\
{\tt CasimirValues} & When set to {\tt True}, substitutes the values of the colour factors corresponding to SU(3). Default option is {\tt False}.
\\
{\tt nnVectorSimplification} & Uses momentum conservation to simplify the dependence of the squared amplitude on the auxiliary four-vectors {\tt nn}. Default option is {\tt False}.
\\
{\tt OverallFactor} & Allows to introduce an overall factor into the contraction of the two input amplitudes. Default value is {\tt "None"}.
\egfb
\phantomcaption{\label{tab:AmplitudeSquareOptions}}
\end{table}
Finally, the {\tt OverallFactor} option allows the inclusion of additional expressions in the contraction of the two input amplitudes. These expressions may include tensor structures such as Dirac matrices, spinors, or colour matrices, for proper contraction with both the amplitude and its conjugate, as well as numerical factors like spin- and color-averaging terms. 

As an example, consider the tree-level computation of the process $g g \rightarrow gg$. The sum over initial spin states, color averaging, and the symmetry factor for identical final-state particles are incorporated using the {\tt OverallFactor} option. Additionally, the auxiliary momentum vectors {\tt nn} arising from the polarization sum of the vector bosons are simplified using the {\tt nnVectorSimplification} option, which expresses the results in terms of Mandelstam variables. The numerical values of the Casimir operators in QCD with $N_c=3$ are substituted using the {\tt CasimirValues} option: 
\exbox{
\inline{1}{& {\tt Amp = GenerateAmplitude[\{G, G\} -> \{G, G\}, Dimension -> 4];} \\ 
}
\inline{2}{& {\tt AmpC = GenerateAmplitudeConjugate[\{G, G\} -> \{G, G\},} \\
   &{\tt  Dimension -> 4];}
}
\inline{3}{& {\tt AmpSq = GenerateAmplitudeSquare[Amp, AmpC,}  \\
   &{\tt nnVectorSimplification -> True, CasimirValues -> True,} \\
  & {\tt OverallFactor -> 1/(2\pw2*8\pw2*2), }\\
   &{\tt  Dimension -> 4];}
}}
\section{From amplitudes to scalar form factors}
\label{sec:form_amps}
%%%%%%%%%%%%%%%%%%%%%%%%%%%%%%%%%%
\subsection{Projectors and form factors}
%%%%%%%%%%%%%%%%%%%%%%%%%%%%%%%%%%
\label{sec:ampProj}
After the Feynman diagrams for a given process have been generated, one is typically interested in amputating the external polarisation vectors and spinors and contracting the external colour and spin indices with some projectors to obtain a set of scalar form factors (see section~\ref{sec:intro}). \anatar\ allows the user to define projectors $P_{j,{\boldsymbol \mu},a}^{\bf rs}$ and to apply them to individual Feynman diagrams as in eq.~\eqref{eq:FF_def} to obtain scalar expressions. We now describe how to perform these manipulations.

The first step is to define the projectors in \anatar. This is achieved by the command:
\newline\newline
{\tt DefineProjector[} \emph{name}{\tt ,} \emph{spins}{\tt ,} \emph{colours}{\tt ,} $N${\tt ,} $L_{\boldsymbol \mu}^{\bf rs}${\tt ,} $C^{\bf a}${\tt ,} \emph{options} {\tt]}
\newline\newline
Here, \emph{name} is simply a \mathematica\ symbol used to refer to the projector. The entry \emph{options} refers to the options {\tt SubstitutionRules} and {\tt CasimirValues}, already discussed in Table \ref{tab:AmplitudeSquareOptions}. The information about a projector is stored inside an association, which can be called via \mbox{{\tt Projector[} \emph{name} {\tt ]}}. A list of all available projectors can be assessed through the variable {\tt AllProjectors[]}. $N$, $L_{\boldsymbol \mu}^{\bf rs}$ and $C^{\bf a}$ are respectively a scalar, a Lorentz/spin tensor and a colour tensor, from which the projector $P_{j,{\boldsymbol \mu},a}^{\bf rs}$ is defined as\footnote{Note that a general projector is a linear combination of such terms. We can always recover any tensor from tensors where the colour and spin indices are factorised from each other.}
\beq
P_{j,{\boldsymbol\mu},a}^{\bf rs} = N\, L_{\boldsymbol\mu}^{\bf rs}\,C^{\bf a}\,.
\eeq
The expressions for the $L_{\boldsymbol\mu}^{\bf rs}\,C^{\bf a}$ can be built using the objects presented in Table~\ref{tab:objectsAmp}. In this process it is of course important to know the type and the range of the indices carried by the tensor.
 The convention for the spin and Lorentz indices is as follows: the number associated with a given index/momentum indicates the position of its respective field in the list \emph{spins}, which is a list of half-integers that defines the spins of the external states. A similar convention holds for the list \emph{colours}, which defines the dimensions of the SU$(3)$ representations. If a tensor carries no Lorentz/spin or colour indices, then the corresponding list is empty, and the corresponding tensor is {\tt 1}. Alternatively, one may use the command
 \newline\newline
{\tt DefineLorentzProjector[ }\emph{name}{\tt ,} \emph{spins}{\tt ,} $L_{\boldsymbol \mu}^{\bf rs}$ {\tt ]}\\
{\tt DefineColourProjector[} \emph{name}{\tt ,} \emph{colours}{\tt ,} $C^{\bf a}$ {\tt ]}\newline\newline
which are convenient for the definition of projectors acting only on the Lorentz and spin indices or the color indices.

Once one or more projectors have been defined by the user, they can be applied to the amplitudes. For example, if the user has defined projectors name {\tt Proj1}, {\tt Proj2}, $\ldots$, he or she may issue the command
\newline\newline
{\tt{pamp}} = {\tt ProjectAmplitude[ amp, \{Proj1,Proj2,...\},} \emph{options} {\tt]}
\newline\newline
where {\tt amp} corresponds to an association output by {\tt GenerateAmplitude[]}. \anatar\ then applies the projectors to each diagram stored in the {\tt Amplitudes} key of {\tt amp}. The result is a new association list, essentially identical to {\tt amp}, but where the value of the {\tt Amplitudes} key now contains the diagrams contracted with the projectors. Note that all functions defined in the previous section to act on the associations produced by {\tt GenerateAmplitude} can also be applied to associations produced by {\tt ProjectAmplitude}. In addition, if {\tt pamp} denotes the association produced by  {\tt ProjectAmplitude}, the function 
\newline\newline
{\tt FormFactor[ pamp, Proj1 ]}
\newline\newline 
allows the user to extract the analytic expression for the form factor after acting with the projector {\tt Proj1}.

Projectors are independent of any given process, and only depend on the external particles. It is therefore not uncommon that the same set of projectors apply to different amplitudes and/or at different loop orders. \anatar\ comes with a set of predefined projectors, which are collected in the file {\tt ProjectorsLib.m}. This file is loaded together with the package and the projectors are available immediately after the package is loaded. The user may extend the list of projectors defined, and they will then be automatically be defined when loading \anatar. The collection of projectors defined in this file can be extended by the user. In order to see which projectors are available that match a given list of \emph{spins} the user may issue
\newline\newline
{\tt ProjectorsBySpins[ }\emph{spins} {\tt ]}
\newline\newline
This returns the list of defined projectors for a set of external particles with a given set of \emph{spins}. The order of the list spins must follow the same order of the fields in the {\tt Process} key of the amplitude. The function {\tt ProjectorsByColours[ }\emph{colours} {\tt ]} works similarly.

As an example, let us consider the QCD corrections to the heavy-quark propagator. Corrections to the self-energy can be organized as
\beq\label{eq:prop}
\Sigma(p_1) = \slashed p_1 \Sigma_1(p_1^2,m) - m\, \Sigma_2(p_1^2,m),
\eeq
where the $\Sigma_i$ are scalar form factors given in terms of the momentum $p_1^{\mu}$ of the fermion and its bare mass $m$. With the SM loaded, the amplitude containing these corrections can be obtained in \ant~via the procedure 

\exbox{
\inline{1}{& {\tt M1=GenerateAmplitude[\{t\}->\{t\},Loops->1,QGrafOptions->"onepi",} \\
 & {\tt CouplingsOrder->\{\{gs,0,2\},\{ee,0\},\{ymt,0\}\},}  \\
 & {\tt PolarizationTrim->True];}
    }
  }
% }
We notice that only one-particle irreducible diagrams are generated. To ensure that no electroweak diagrams are generated, the option {\tt CouplingsOrder} is included. In addition, the option {\tt PolarizationTrim} is set to {\tt True}, as should be done with every generated amplitude that subsequently is contracted with projectors. Thus, a single diagram is produced. We can check the available projectors defined in {\tt ProjectorsLib.m} associated with external fermions by issuing
\exbox{
\inline{2}{& {\tt  ProjectorsBySpins[\{1/2,1/2\}]}}\\
\breakline\\
\outline{2}{& {\tt \{projTT1,projTT2\}}\\ 
    }
  }
Similarly, we can access the available projectors based on their SU$(N_c)$ color representations, i.e., the fundamental and adjoint representations. 
\exbox{
\inline{3}{& {\tt  ProjectorsByColours[\{3,3\}]}}\\
\breakline\\
\outline{3}{& {\tt \{projTT1,projTT2\}}\\ 
    }
  }
The information of these projectors can be accessed as
\exbox{
\inline{3}{& {\tt Projector[projTT1]}}\\
\breakline\\
\outline{3}{& {\tt <|Name->projTT1,}\\ 
 & {\tt \quad Spins->\{1/2,1/2\},}\\
 & {\tt \quad ColourRepresentation->\{3,3\},}\\
 & {\tt \quad PCoefficient->1/(4*p1.p2),}\\
 & {\tt \quad PLorentz->GammaM[p1, Spin1, Spin2],}\\
 & {\tt \quad PColour->Metric[Colour1, Colour2],}\\
 & {\tt \quad Operator->(Metric[Colour1,Colour2]*}\\
 & {\tt \quad\quad\qquad\qquad GammaM[p1, Spin1, Spin2])/(4*p1.p2)|>}\\
    }
  }
We note that the operator that acts on the scattering amplitudes, under the tag {\tt Operator}, is the product of the entries in {\tt PCoefficients}, {\tt PLorentz} and {\tt PColor}. In addition, it can be observed that the indices {\tt Spin1} and {\tt Colour1}, and the momentum {\tt p1} correspond to the first top quark in the list {\tt \{t,t\}}. 

Let us now compute the form factor $\Sigma_i$ from eq.~\eqref{eq:prop}. This means that in the case of the top self-energy, there are two projectors. The amplitudes $\Sigma_i$ are then obtained as
\exbox{
\inline{4}{& {\tt PM1=AmplitudesSimplify[ProjectAmplitude[M1,\{projTT1,projTT2\}]]} \\
 }\\
\breakline\\
\outline{4}{& {\tt <|Process->\{\{t[p1]\} -> \{t[p2]\}\}, } \\
 & {\tt \; Total->\{1;;1\},} \\
 & {\tt \; Name->"t\_t",} \\
 & {\tt \; LoopOrder->1,}\\
 & {\tt \; Projectors->2,}\\
 & {\tt \; Kinematics->\{p1\pw2->S,p1.p1->S,(p1.p1)\pw(-1)->S\pw(-1),p2->p1\},}\\
 & {\tt \; Amplitudes->}  \\
 & {\tt \; \{\{DiagramID[1],projTT1\}->}  \\
 & {\tt \qquad\qquad(CF*(d-2)*GC10\pw2*Nc*Den[k1,MT]*Den[k1-p1,0]*p1.k1)/S, } \\
 & {\tt \;\quad\{DiagramID[1],projTT2\}-> }\\
 & {\tt \qquad\qquad-(CF*d*GC10\pw2*Nc*Den[k1,MT]*Den[k1-p1,0])\}|> }
 }
}
where $S=p_1^2$ and the {\tt CA}, {\tt CF} and {\tt Nc} are the quadratic Casimir eigenvalues of $SU(N_c)$ gauge group. We apply the function {\tt AmplitudesSimplify} to the individual form factors to obtain compact expressions. The result of the application of the projector {\tt projTT}$x$ acting on the amplitude corresponding to the $i^{\mathrm{th}}$ diagram is labelled as {\tt \{DiagramID[i],projTT}$x${\tt\}}. In the example above, since we have one diagram and two projectors, we have {\tt \{DiagramID[1], projTT1\}} and {\tt \{DiagramID[2], projTT2\}}.  Consequently, the first and second elements of the list correspond to the contributions to $\Sigma_1$ and $\Sigma_2$ that arise from the single diagram produced.

\subsection{Mapping to integral families}
\label{sec:topology_mapping}

After projection to form factors, amplitudes are linear combinations of scalar Feynman integrals, cf. eq.~\eqref{eq:workflow_decomp}. The latter can be sorted into \emph{integral families} or \emph{topologies}, and for each integral family we may identify a basis. In this section we discuss how we can use \anatar\ to identify the topologies for a given process. The reduction to a basis of master integrals will be addressed in section~\ref{sec:IBP}.

The first step consists in identifying the integral families that need to be considered. Often the relevant integral families are already available in the literature (and the corresponding basis of master integrals was identified, and possibly even evaluated). The user can define integral families via the command
%
%The procedure starts with the definition of the integral families, which in \ant~is done by means of the command
\newline\newline
{\tt DefineTopology[}\emph{name}{\tt , }\emph{loopMomenta}{\tt, }\emph{extMomenta}{\tt ,}\emph{ denominators}{\tt]}
\newline\newline
where \emph{name} is a \mathematica~symbol which labels the topology. 
The list \emph{denominators} represents the set of denominators $D_p$ that define the integral family.\footnote{Often additional requirements are imposed, like that every scalar product involving a loop momentum can be expressed in terms of denominator. At this stage, no such requirement is made.}  The topology information is saved in association form and can be accessed via {\tt Topology[}\emph{name}{\tt ]}. The {\tt Propagators} key lists the denominators $D_p$ that define the topology. In addition, the list of all the topologies defined in the current \mathematica~kernel can be obtained by issuing the command {\tt AllTopologies[]}.

Alternatively, the user may prompt \ant~to identify the topologies needed to classify all the scalar integral automatically. 
%Assume that {\tt pamp} is the association produced by the {\tt ProjectAmplitude[]} function. 
The user can instruct \anatar\ to identify a set of topologies for this process by issuing
\newline\newline
{\tt FindTopology[pamp, } \emph{options}{\tt]}\,.
\newline\newline
The execution of this function returns, and automatically defines, the necessary integral families to fully map the amplitudes. Internally, {\tt FindTopology} extracts the list of propagators from all the Feynman diagrams and then searches for symmetries between any two of them. These symmetries arise from the general properties of $D$-dimensional integration, such as linearity, translational invariance, and scaling behaviour. To achieve this, we start with a Feynman diagram A, we then identify all possible linear relations that preserve the invariance of graph A with respect to translational invariance in terms of the external momenta. Once these relations are determined, we substitute them into the propagator set of another Feynman diagram, namely graph B. If the substitution reveals that graphs A and B are related through these symmetries, we classify them as having the same topology. If no such relation exists, graph B is assigned to a new integral family. This process is repeated for all Feynman diagrams to identify the independent set of topologies. It is possible that the selected topology obtained from analyzing all Feynman diagrams is not sufficient to form a complete basis for spanning all scalar products. In such cases, the user must manually add additional denominators ({auxillary propagators or ISPs}) as needed to complete the basis. The function {\tt FindTopology[]} has a single option that specifies whether or not to automatically define the topologies found in the mapping of the amplitudes. More precisely, when the option {\tt DefineTopology} is set to {\tt False}, the denominators of the resulting topologies are output as a list, but they are not automatically defined using the {\tt DefineTopology[]} function. The user may still define them manually at a later stage.

After all topologies have been identified and defined which spans the set of all scalar products, the user may instruct \anatar\ to map all scalar from factors to the individual topologies by issuing
\newline\newline
{\tt AmplitudeToTopologies[pamp, }\emph{options}{\tt]}\,,
\newline\newline
where again {\tt pamp} is the association produced by {\tt ProjectAmplitude[]}.
The output of {\tt AmplitudeToTopologies[]} is an association identical to the input {\tt pamp}, but with the expressions in the {\tt Amplitudes} key now written in terms of Feynman integrals sorted according to the integral families. This implies that the function {\tt AmplitudeToTopologies[]} also writes every scalar product in the numerator that involves loop momenta in terms of the denominators $D_p$. For this of course it is mandatory that the topologies defined by the user satisfy this property. The notation in \ant~for the integrals $I^{F_i}_{\boldsymbol{\nu}}$ in eq.~\eqref{eq:ScalarFI} is
\newline\newline
$I^{F_i}_{\nu_1,\ldots,\nu_p} \longrightarrow $ {\tt TopInt[}\emph{name}{\tt , }$\nu_1${\tt \,,..., }$\nu_p${\tt ]},
\newline\newline
where \emph{name} stands for the label $F_i$ and the sequence $\nu_1${\tt \,,..., }$\nu_p$ corresponds to the exponents of the denominators $D_p$ in the order in which they appear in the definition of the topology. By default, \anatar\ attempts to map the scalar integrals to any of the topologies defined in the list {\tt AllTopologies[]} (if more than one mapping is possible, only the first successful mapping is retained). 
In case the user wants to use a subset of all the defined topologies, he or she may do so using the option {\tt Topologies}, which follows the syntax follows
\newline\newline
{\tt Topologies->\{Topo1,Topo2,...\}}
\newline\newline

Let us illustrate this on the example of the heavy-quark self-energy from section \ref{sec:ampProj}. It is enough to define the single bubble topology with one massive and one massless denominator. This is done by
\exbox{
\inline{5}{& {\tt DefineTopology[BubbleTT, \{k1\}, \{p1\}, \{Den[k1,MT],Den[k1-p1,0]\}]}}
}

The amplitudes obtained after the actions of projectors saved in {\tt PM1}, can be mapped into the above topology as follows: 
\exbox{
\inline{6}{& {\tt TM1=AmplitudesSimplify[AmplitudeToTopologies[PM1]];}  }\\
\inline{7}{& {\tt TM1[Amplitudes]} } \\
 %& {\tt \qquad\qquad\qquad\qquad\qquad\qquad\qquad \{projTT1,projTT2\}]]}
\breakline\\
\outline{7}{& {\tt \; \{\{DiagramID[1],projTT1\}->} \\
  & {\tt \qquad\qquad (CF*(d-2)*GC10\pw2*Nc*(TopInt[BubbleTT,0,1]+}\\
 & {\tt \qquad\qquad (S + MT\pw2)*TopInt[BubbleTT,1,1]-}\\
 & {\tt \qquad\qquad TopInt[BubbleTT,1,0]))/(2*S),}\\
 & {\tt \;\; \{DiagramID[1],projTT2\}->}\\
 & {\tt \qquad\qquad -(CF*d*GC10\pw2*Nc*TopInt[BubbleTT, 1, 1])\} }  
 }
}
We only show the expressions listed in the {\tt Amplitudes} key, given that they bear the only changes introduced by {\tt AmplitudeToTopologies}. These expressions are now written in terms of objects of the form {\tt TopInt[BubbleTT,...]}, which represent the integrals written in terms of the topology {\tt BubbleTT} defined above. Notice that, as a cross-check, these analytic expressions should not contain any explicit dependence on the loop momenta, not even through objects of the form {\tt Den[...]}.

%%%%%%%%%%%%%%%%%%%%%%%%%%%%%%%%%%%%%%%%%%%
\subsection{Reduction to master integrals}
%%%%%%%%%%%%%%%%%%%%%%%%%%%%%%%%%%%%%%%%%%%
\label{sec:IBP}

After all integrals have been mapped to topologies, one is typically interested in expressing all the members of a given family in terms of a minimal set of basis integrals, called \emph{master integrals} (MIs). There are various strategies to reduce a given member of an integral family to a basis. The most prominent approach is the so-called \emph{Laporta algorithm}~\cite{Laporta:2000dsw} based on \emph{integration by parts (IBP)} identities~\cite{Chetyrkin:1981qh,Tkachov:1981wb}, 
and there are several public codes to perform the IBP reduction, cf.,~e.g.,~ refs.~\cite{Anastasiou:2004vj,Smirnov:2008iw,Studerus:2009ye,vonManteuffel:2012np,Lee:2012cn,Lee:2013mka,Maierhofer:2017gsa,Klappert:2020nbg,Smirnov:2023yhb,Wu:2025aeg}. Each of these codes takes as an input the definitions of the topologies as well as the list of integrals that appear in the form factors. Hence, after the procedure described in section~\ref{sec:topology_mapping}, \anatar\ has all the information needed to run these codes. However, each of these codes requires the information to be defined in a format specific to each code. One may write dedicated interfaces for each code, and \anatar\ comes equipped with two such interfaces to the publicly-available IBP reduction codes \kira~\cite{Klappert:2020nbg,Maierhofer:2017gsa} and \litered~\cite{Lee:2012cn,Lee:2013mka}.\footnote{Currently only \litered\ version 2 is supported. If an earlier \litered\ version \anatar\ is called from within \anatar, an error message is issued.} Before we describe the usage of these interfaces, we have to make a comment. Performing the IBP reduction for state-of-the-art multi-scale multi-loop computations can be extremely challenging, and a lot of effort has gone over the last years in improving the IBP reduction, cf.,~e.g.,~refs.~
\cite{Laporta:2000dsw,Smirnov:2005ky,Smirnov:2006tz,Lee:2008tj,Schabinger:2011dz,vonManteuffel:2014ixa,Georgoudis:2016wff,Peraro:2016wsq,Peraro:2019svx,Barakat:2022qlc,Wu:2023upw,Mokrov:2023vva,Belitsky:2024jhe,Smirnov:2024onl,Chestnov:2024mnw,Wu:2025aeg}, or by developing methods that would allow one to circumvent the solution of large linear systems altogether~\cite{Mastrolia:2018uzb,Frellesvig:2019kgj,Frellesvig:2019uqt,Frellesvig:2020qot,Weinzierl:2020xyy,Caron-Huot:2021xqj,Caron-Huot:2021iev,Chestnov:2022xsy,Brunello:2023rpq,Fontana:2023amt,Brunello:2024tqf}. Here we do not intend to address the issue of efficiency of feasibility of the IBP reductions. \anatar\ merely provides a means to interface to exiting tools for IBP reduction. Whether or not the reduction can be achieved depends on the amplitude considered and on the tool to which \anatar\ is interfaced.

\subsubsection{Interface to \litered}
\label{sec:litered}
Let us start by discussing the interface to \litered. \litered\ is written in \mathematica, and \anatar\ has been designed so that it is possible to have both package loaded simultaneously into the same \mathematica\ session (provided that \litered\ is loaded first\footnote{\litered\ relies on \fermatica~\cite{fermatica}, and also \fermatica\ can be loaded into the same \mathematica\ session as \anatar.}). Since \litered\ and \anatar\ can be loaded simultaneously, the user can just use \litered\ in the standard way. In addition, \anatar\ provides a command to define and reduce the topologies in \litered. This can be done via
\newline\newline
{\tt RunLiteRed[}\emph{topoList}{\tt ,} \emph{options}{\tt ]}
\newline\newline
where \emph{topoList} is the list of the name of the topologies that the user wishes to reduce. If this variable is set to {\tt All}, then the list {\tt AllTopologies[]} of all defined topologies is used instead. \anatar\ will then write in the process folder a \mathematica\ file for each topology that contains all the topology definitions needed to run \litered. \emph{options} is a list of optional arguments, mostly related to additional optional arguments related to the running of \litered. For a complete list of options, see Table~\ref{tab:litered_options}. The resulting file is then loaded and executed in \mathematica, prompting \litered\ to perform the reduction. It is also possible to just output the input file for \litered\ without executing it by issuing {\tt WriteLiteRedFile[]} instead of {\tt RunLiteRed[]}. The associations representing the topologies are updated to include the keys {\tt LiteRedFile} and {\tt LiteRedMIs}, which refer to the input file for \litered\ and the list of master integrals obtained, respectively. It is possible to  delete the file produced by \anatar\ and the keys {\tt LiteRedFile} and {\tt LiteRedMIs} from the topologies {\emph{topo1}, \emph{topo2}, $\ldots$} by issuing 
\newline\newline
{\tt ClearLiteRed[}\emph{topo1}, \emph{topo2}, $\ldots${\tt]}
\newline\newline
If the argument list is empty, then {\tt ClearLiteRed[]} is executed for every topology defined in {\tt AllTopologies[]}.

The result of the reduction by \litered\ can be directly used within \anatar. In particular, it is possible to instruct \anatar\ to insert the reduction to MIs directly into the {\tt Amplitudes} key of the associations that represent the amplitudes mapped to topologies. For example, if {\tt Tamp} represents the output of the function {\tt AmplitudeToTopologies[]}, then we can insert the result of the IBP reduction with \litered by issuing:
\newline\newline
{\tt LiteRed[IBPReduce][Tamp]}
\newline\newline
More generally, it is possible to act with a function {\tt func[]} provided by \litered\ directly on the associations produced by \anatar\ by wrapping it into the {\tt LiteRed[]} container,
\newline\newline
{\tt LiteRed[func][Tamp]}
\newline\newline

Let us again illustrate this on the example of the heavy-quark self-energy. We may reduce the topology {\tt BubbleTT} from the previous section with \litered\ by issuing
\exbox{
\inline{8}{& {\tt RunLiteRed[BubbleTT];}}\\
\inline{9}{& {\tt LRM1=AmplitudesSimplify[LiteRed[IBPReduce][TM1]];}  }\\
\inline{10}{& {\tt LRM1[Amplitudes]} } \\
 %& {\tt \qquad\qquad\qquad\qquad\qquad\qquad\qquad \{projTT1,projTT2\}]]}
\breakline\\
\outline{10}{& {\tt \; \{\{DiagramID[1],projTT1\}->} \\
 & {\tt \qquad\quad (CF*(d-2)*GC10\pw2*Nc*((S + MT\pw2)*TopInt[BubbleTT, 1, 1]-}\\
 & {\tt \qquad\quad TopInt[BubbleTT, 1, 0]))/(2*S),}\\
 & {\tt \;\; \{DiagramID[1],projTT2\}->}\\
 & {\tt \qquad\quad -(CF*d*GC10\pw2*Nc*TopInt[BubbleTT, 1, 1])\} }  
 }
}
The topology {\tt BubbleTT} has been reduced, so that the resulting expressions are then given in terms of a bubble and a tadpole integral.
\begin{table}
\bgfb
\multicolumn{2}{c}{\textbf{Table~\ref{tab:litered_options}: Options of {\tt RunLiteRed}/{\tt WriteLiteRedFile}} }\\
\\
{\tt Save} & If {\tt True}, the result of the \litered\ reduction is saved on file.
\\
{\tt Directory} & The name of the directory where the result of the \litered\ reduction should be saved.
\\
{\tt Scalars} & The names of the scalar variables to be declared for \litered.\\
{\tt Vectors} & The names of the vector variables to be declared for \litered, other than the loop and external momenta that appear in the definition of the topology.\\
{\tt LiteRedOptions} & The list of rules which should be passed as options to the function {\tt NewDsBasis[]} in \litered. The default is {\tt SolvejSector -> True}.\\
{\tt ClearLiteRed} & Remove all the information about LiteRed files produced by \anatar\ from a topology.
\egfb
\phantomcaption{\label{tab:litered_options}}
\end{table}

\subsubsection{Interface to \kira}
\label{sec:kira}

We now describe the interface to \kira.\footnote{Currently only \kira\ version 2 is supported. } Several files are required for the successful IBP reduction of the integral families. \ant~is designed to generate those files, execute them, and import back the reductions of the loop integrals present in the amplitudes. In order to obtain form factors in terms of master integrals it is enough to issue
\newline\newline
{\tt IntegralReduceKira[ Tamp,} \emph{options}{\tt ]}
\newline\newline
where {\tt Tamp} refers to the output of the function {\tt AmplitudesToTopologies[]}. The set of \emph{options} can be found in Table \ref{tab:OptionsKira}. We notice that topologies do not need to be provided again, since they have already been defined for the mapping of the amplitudes performed by {\tt AmplitudesToTopologies}. The output of {\tt IntegralReduceKira[]} is an association with exactly the same structure as the input {\tt Tamp}, but with the expressions in the {\tt Amplitudes} key now expressed in terms of a reduced set of integrals. This set corresponds to the MIs and is stored in the list {\tt MasterIntegrals[]}. To generate this list, \ant~executes internally the function {\tt IdentifyIntegrals[expr]}, which produces a list with all loop integrals mapped to any defined topology present in the expression {\tt expr}. We also notice that running {\tt IntegralReduceKira[]} additionally creates a folder named {\tt kira\_reduction}$x${\tt L} located in the process folder, with $x$ the number of loops. There, the user can find all the files that are usually associated with the execution of \kira.

\begin{table}
\bgfb
\multicolumn{2}{c}{\textbf{Table~\ref{tab:OptionsKira}: Options of {\tt IntegralReduceKira} }}\\
\\
{\tt Topologies} & Selects the topologies to be reduced. The default setting is to consider all of the topologies in {\tt AllTopologies[]}.
\\
{\tt BasisMI} & Defines the preferred basis of master integrals. Default value is {\tt Automatic}, thus allowing \kira\ to choose the basis.
\\
{\tt KiraSectors} & Selects a sector for a given topology. Default value is {\tt Automatic}.
\\
{\tt SeedSelection} & Default value is {\tt Default}, which means that $r$ and $s$ are obtained from the loop integrals appearing in the amplitude.
\\
{\tt OnlyWriteFiles} & When set to {\tt True}, writes all the required input files by \kira~without executing them.
\\
{\tt OnlyImport} & When set to {\tt True}, imports the IBP reduction of the integrals in the amplitude without running \kira. The user can also provide the path to the file with the reduction. The convention for the name of the topologies and symbols should follow the same of the input amplitude. Default is {\tt False}.
\\
{\tt CutPropagators} & In cases where the user needs to specify the cut propagators manually, this option allows them to do so by providing the propagators as a list.
\\
\egfb
\phantomcaption{\label{tab:OptionsKira}}
\end{table}
In analogy to the reduction obtained with \litered~in the previous section, the reduction of the topology, and thus the form factors, in the example of the heavy-quark self-energy is achieved by executing the command
\exbox{
\inline{7}{& {\tt KRM1=IntegralReduceKira[TM1];}  }\\
}
We omit the output here, as it is identical to that obtained in in the output {\tt LRM1}.  

Notice that among the options listed in Table \ref{tab:OptionsKira} we have the option {\tt SeedSelection}. This allows the user to manually to select the values of the parameters $r$ and $s$, defined as
\begin{align}
    r = \sum_{j=1}^p \nu_j\,\theta\Big(\nu_j - \frac{1}{2}\Big), \qquad&\qquad s = -\sum_{j=1}^p \nu_j\,\theta\Big( \frac{1}{2} - \nu_j \Big).
\end{align}
Thus, $r$ and $s$ represent the sum of all positive powers and the negative sum of all negative powers, respectively. These values are automatically obtained by \ant~from the list of integrals that appear in the amplitudes. \kira~generates equations only for sets $\boldsymbol{\nu}$ for which $r \leq r_{max}$ and $s \leq s_{max}$. The manual setting of these parameters can be done by adding the option
\newline\newline
{\tt SeedSelection -> \{ r, s \} }
\newline\newline
The values of {\tt r} and {\tt s} above can be integer numbers or strings. The latter should be used in the case the syntax of \kira~for ranges of the type $[r_{min},r_{max}]$ is used. In some cases, the selection of $r$ and $s$ could be useful to find an optimal basis of MIs. In other cases, it might happen that, when $s$ is chosen too small, \kira~does not find a relation between the integrals in the amplitude and a given preferred basis of MIs provided in the option {\tt BasisMI}. In such cases, varying the parameters $r$ and $s$ could lead to the desired result. Observe that setting a large $s$ parameter could be computationally intense.

The package also provides a scope to include cut propagators for processes involving real emissions. This must be specified during the \kira~ reduction, and the propagators can be supplied using the {\tt Den} function of our package as follows:
\newline\newline
{\tt CutPropagators -> \{ Den[a1, b1], Den[a2,b2],...\} }
\newline\newline

The user can also select sectors for the reduction performed by \kira. This can be done by adding the option
\newline\newline
{\tt KiraSectors -> \{\{Topo1,sec1\},\{Topo2,sec2\},...\} } 
\newline\newline
The reduction of {\tt Topo}$x$ will then be restricted to the specified sector {\tt sector}$x$. The syntax above implies that it is not possible to request the reduction of more than one sector per topology  in \anatar, nor is it not possible to select individual values of $r$ and $s$ for each of these reductions. In such specific cases, the user should run \kira~manually. For more information, we refer the user to the \kira~documentation.

\section{Validation}
\label{sec:validation}
% !TEX root = anatar.tex
\renewcommand{\arraystretch}{1.2}
\setlength{\extrarowheight}{1pt}
\begin{table}
\small
    \centering
    \begin{tabular}{|c|c|c|c|}
        \hline
        \makecell{Perturbative \\ Order}  &  Process  &  Error &  Reference  \\ 
        \hline \hline

        \multirow{18.4}{*}{\centering Tree-Level} 
        & $G G \to G G$                                     & $1.59044\times 10^{-7}$ &  \mg \\
        & $b G \to H b$                                     & $\num{1.099805174}\times 10^{-7}$ &  \mg \\
        & $d \bar{d} \to d \bar{d}$             & - &  \formC \\
        & $d \bar{d} \to t \bar{t}$               & - &  \formC \\
        & $d \bar{d} \to Z G$                               & $\num{6.39536}\times10^{-6}$ &  \mg \\
        & $u\bar{u} \to W^+W^-$                             & - & \formC \\
        & $u\bar{u} \to ZZ$                                 & $5.18743\times10^{-10}$ & \mg \\
        & $u\bar{u} \to ZA$                                 & $2.18669\times 10^{-7}$ & \mg \\
        & $u\bar{u} \to AA$                                 & $9.80525\times10^{-8}$ & \mg \\
        & $u\bar{u} \to s\bar{s}$                           & - & \formC \\
        & $u\bar{d} \to c\bar{s}$                           &$2.26711\times 10^{-7}$  & \mg \\
        & $us \to cd$                                       & $9.28108\times 10^{-8}$ & \mg \\
        & $t \bar{t} \to t \bar{t}$               & -  & \formC \\
        & $G G \to t \bar{t}$                               & $1.51196\times10^{-7}$ &  \mg \\
        & $t\bar{t} \to ZA $                                & $3.69093\times10^{-7}$  &  \mg \\
        & $t\bar{t} \to AA$                                 & $7.3066\times10^{-7}$ &  \mg \\ 
        & $t\bar{t} \to u\bar{u}$                           & $2.70361\times 10^{-7}$ & \mg \\ 
        % & $t\bar{t} \to W^+W^-$   & $0.0000336612$ &  \mg \\ 
        & $t\bar{t} \to W^+W^-$                             & - &  \formC \\ 
        & $t\bar{t} \to ZZ$                                 & - &  \formC \\ [1.5ex]
        
        \hline \hline

        \multirow{2.3}{*}{\centering One-Loop} 
        & $q \bar{q}  \to Z $                    & - &  \cite{Harlander:2003ai}\\
        & $q \bar{q}  \to Z G$                   & - &  \cite{Harlander:2003ai}\\ [1.5ex]

        \hline \hline

        \multirow{6.3}{*}{\centering Two-Loop} 
        & $t \to t $                             & - & \cite{Tarasov:1996bz} \\
        & $G \to G $                             & - &  \cite{Duhr:2025zqw} \\
        & $G G \to H $                             & - & \cite{Anastasiou2007} \\
        & $b \bar{b}  \to H $                    & - & \cite{Harlander:2003ai} \\
        & $b \bar{b}  \to H G$                   & - &  \cite{Harlander:2003ai}\\
        & $b \bar{b}  \to H G G$                 & - &  \cite{Harlander:2003ai}\\ [1.5ex]
        \hline
    \end{tabular}
    \caption{Selection of quark and gluon processes used for validation of the SM in \ant. The dash in the error entry stands for those processes that have been validated analytically, and thus the expressions match identically. }
    \label{tab:checks1}
\end{table}

In order to validate our code, we have compared the output of \anatar\ with results available in the literature and/or obtained by other codes (either numerically or analytically). 
Specifically, we have compared analytic expressions obtained at tree-level with \ant~ with well-established amplitude generation codes such as \formC~\cite{Hahn:2016ebn}~and \mg~\cite{Alwall2014}. Amplitudes obtained at higher-orders in \ant~have been validated with results from the literature. 

At tree-level, checks have been performed  for unintegrated squared matrix elements in Feynman gauge, for both the QCD and electroweak sectors. The checks with \mg~were performed choosing a benchmark point in the phase space, and we observe very good numerical agreement. We also verified several analytic results using \formC. We checked a total of 36 key processes, which are shown in Tables \ref{tab:checks1} and \ref{tab:checks2}. In some tree-level cases, the cross-check with \mg~produced error estimates of order $~10^{-5}$. In such cases, we also validated the processes analytically using \formC.
At one loop, we checked the virtual contributions to Higgs production through bottom-quark annihilation, as well as the real-emission and double real-emission subprocesses to this particular process. These are compared against available analytical results, up to and  including the stage of the reduction to MIs. 

In addition to the checks listed in Tables \ref{tab:checks1} and \ref{tab:checks2}, we also performed checks for selected higher-order processes with the inclusion of higher-dimensional effective operators. These checks include the gluon-fusion Higgs production with two-loop QCD corrections in the SMEFT \cite{Deutschmann2017} and the QCD corrections to the self-energy of quark and gluon fields at two-loop in the SMEFT \cite{Duhr:2025zqw}.

We note that whenever a process  appears as validated at a given order, it is understood that it has been validated at previous orders. For example, if a process appears as checked at two loop, then it was also checked at tree-level and at one loop. Similarly, if a process has been validated at one loop, it means that it has also been checked at tree-level.

\renewcommand{\arraystretch}{1.2}
\setlength{\extrarowheight}{1pt}
\begin{table}[!t]
\small
    \centering
    \begin{tabular}{|c|c|c|c|}
        \hline
        \makecell{Perturbative \\ Order}  &  Process  &  Error &  Reference  \\ 
        \hline \hline

        \multirow{16.4}{*}{\centering Tree-Level} 
        & $\tau \bar{\tau} \to \tau \bar{\tau}$   & - &  \formC \\
        & $e^+ e^- \to e^+ e^-$                & - &  \formC \\
        & $e^+ e^- \to \tau^+ \tau^-$                       & $8.84025\times10^{-8}$ & \mg  \\
        & $e^+ e^- \to e^+ e^- A$                           &$1.86181\times10^{-8}$  &  \mg \\
        & $e^+ e^- \to \nu_e \bar{\nu}_e$                           & - & \formC \\
        & $e^+ e^- \to t \bar{t}$                           & - & \formC \\
        & $e^+ e^- \to b \bar{b}$                           & - & \formC \\
        & $d \bar{d} \to e^+ e^- $  & - & \formC  \\
        &  $d \bar{d} \to e^+ e^- G$   & - & \formC  \\
        &  $\tau^+\tau^- \to W^+W^-$   & - & \formC  \\
        &  $\tau^+\tau^- \to ZZ$   & - & \formC  \\
        &  $\tau^+\tau^- \to ZA$   & - & \formC  \\
        &  $\tau^+\tau^- \to AA$   & - & \formC  \\
        &  $W^+W^- \to ZZ$   & - & \formC  \\
        &  $ZZ \to ZZ$   & - & \formC  \\
        &  $HH \to ZZ$   & - & \formC  \\
        &  $HH \to W^+W^-$   & - & \formC  \\
        
        \hline 
    \end{tabular}
    \caption{Selection of lepton and weak boson processes used for validation of the SM in \ant. The dash in the error entry stands for those processes that have been validated analytically, and thus the expressions match identically.}
    \label{tab:checks2}
\end{table}

\section{An example calculation: Top self-energy at two loops in the SM}
\label{sec:example}

\begin{figure}[h]
\centering
\unitlength = 0.6mm
    \begin{subfigure}{0.3\textwidth}
    \centering
    \includegraphics[width=0.8\linewidth]{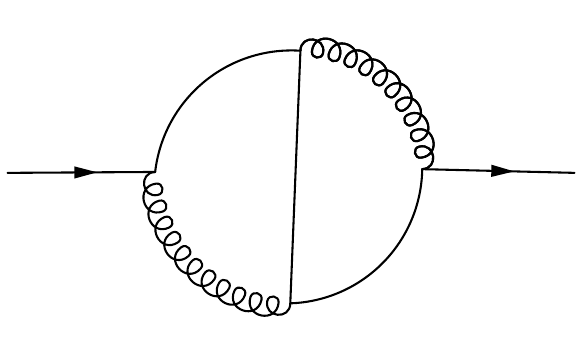}
    %\caption{ }
    \end{subfigure}    
    \hfill
    \begin{subfigure}{0.3\textwidth}
    \centering
    \includegraphics[width=0.8\linewidth]{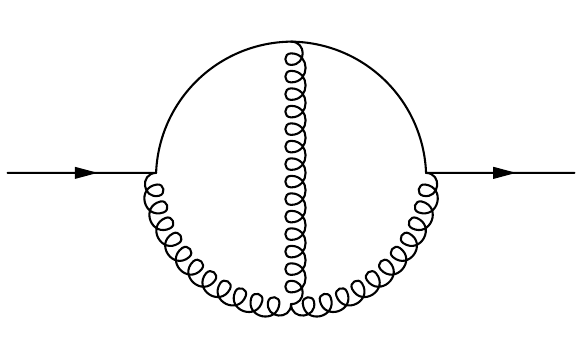}
    %\caption{ }
    \end{subfigure}
    \hfill
    \begin{subfigure}{0.3\textwidth}
    \centering
    \includegraphics[width=0.8\linewidth]{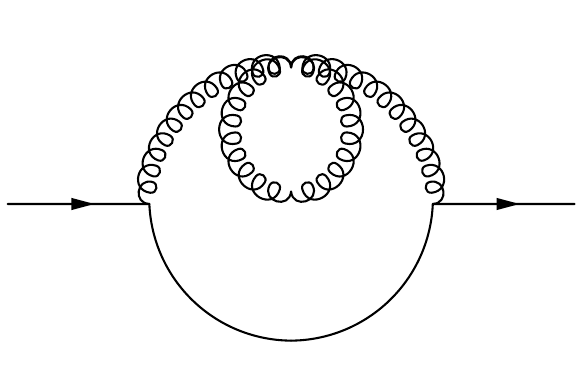}
    %\caption{ }
    \end{subfigure}

\vspace{1em}
    \begin{subfigure}{0.3\textwidth}
    \centering
    \includegraphics[width=0.8\linewidth]{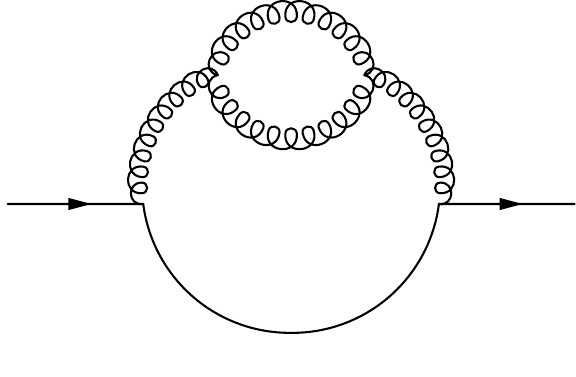}
    %\caption{ }
    \end{subfigure}
    \hfill
    \begin{subfigure}{0.3\textwidth}
    \centering
    \includegraphics[width=0.8\linewidth]{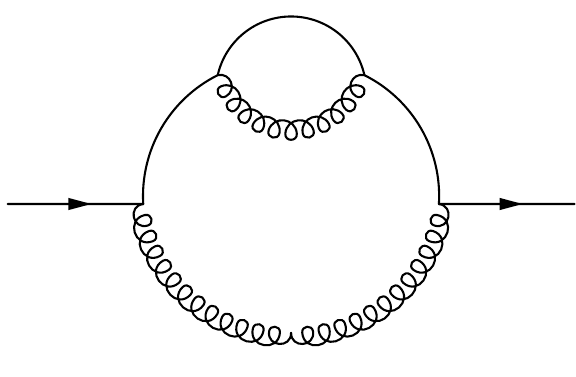}
    %\caption{ }
    \end{subfigure}
    \hfill
    \begin{subfigure}{0.3\textwidth}
    \centering
    \includegraphics[width=0.8\linewidth]{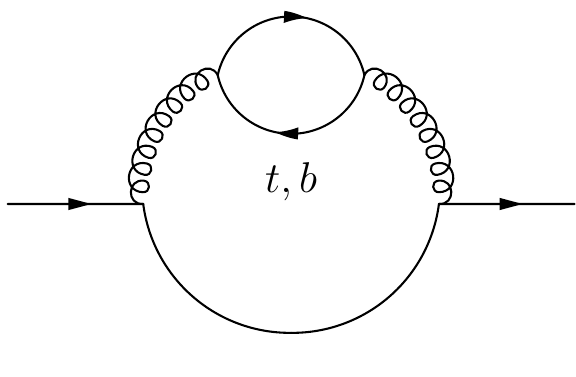}
    %\caption{ }
    \end{subfigure}
    
\vspace{1em}
    \begin{subfigure}{0.3\textwidth}
    \centering
    \includegraphics[width=0.8\linewidth]{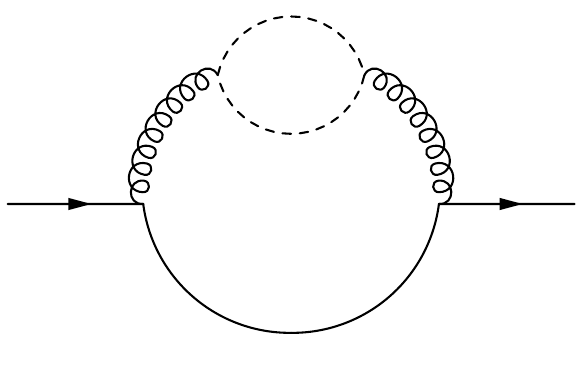}
    %\caption{ }
    \end{subfigure}
    \caption{Feynman diagrams corresponding to the self-energy of the top quark in the SM. The penultimate diagram actually represents two distinct cases, one with the top quark running in the inner fermion loop and one with the bottom quark. The final diagram corresponds to the contribution with the gluon ghost running in the internal loop. }
    \label{fig:top-self}
\end{figure}

In this section, we consider the two-loop QCD corrections to the top self-energy. This is a well-documented computation. The aim is to showcase the capabilities of \ant~when dealing with relevant quantum corrections for phenomenology studies by reproducing the results of ref. \cite{Fleischer1999}. The results are presented in the covariant gauge. We can generate the required amplitudes by issuing
\exbox{
\inline{1}{& {\tt LoadModel["sm"]}}\\
\inline{2}{& {\tt cOrders=\{\{gs,0,4\},\{ee,0\},\{MT,0\}\};}}\\
\inline{3}{& {\tt noPart=\{H,u,d,c,s,G0\};}}\\
\inline{4}{& {\tt gauges=\{"QCD"->"RXi"\};}}\\
\inline{5}{& {\tt M2=GenerateAmplitude[\{t\}->\{t\}, Loops->2,} \\
  & {\tt \qquad QGrafOptions->"onepi", PolarizationTrim->True,} \\
  & {\tt \qquad CouplingsOrder->cOrders, RemovePropagator->noPart,} \\
  & {\tt \qquad  Gauges->gauges];} }\\
\inline{6}{& {\tt PM2=ProjectAmplitude[M2, \{projTT1,projTT2\},} \\
  & {\tt \qquad\qquad\qquad\qquad\qquad\qquad\qquad CasimirValues -> True]/.\{MB->0\};} }
  }
This generates the eight two-loop diagrams shown in fig.~\ref{fig:top-self}. 

The option {\tt Gauges} selects the covariant gauge for the gluon propagator. 
We notice that the light quarks lead to multiple diagrams which are equivalent. For efficiency, only the diagram with the bottom quark is computed. Hence we used the option {\tt RemovePropagators} to remove the other diagrams involving light quarks. Additionally, the mass of the bottom quark is set to zero. From this, to account for the number of light quarks in terms of the total number of flavours ($n_f$), we multiply the form factors in the projected amplitudes corresponding to diagrams with insertions of the bottom quark. To identify these diagrams, we refer to the \qgraf~output and find them by manually drawing the diagrams. In this case there is a single diagram, corresponding to the tag {\tt DiagramID[8]} The function {\tt FindFermionLoop[PM2]} can also be used to find diagrams with an internal closed fermion loops, which in the top self-energy case gives two diagrams (7th and 8th). Then by checking the \qgraf~output one can see that the 7th and 8th diagrams correspond to the top and bottom loops, respectively. Thus, we can redefine such amplitudes by using the function {\tt AmplitudesMap[]} as

\begin{figure}[h]
\centering
    \begin{subfigure}{0.3\textwidth}
    \centering
    \includegraphics[width=1\linewidth]{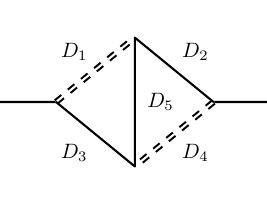}
    \end{subfigure}
\qquad\quad
    \begin{subfigure}{0.3\textwidth}
    \centering
    \includegraphics[width=1\linewidth]{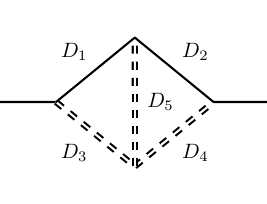}
    \end{subfigure}
    \caption{\label{fig:T_topos}Kite topologies associated to the corrections of the heavy-quark self-energy at two loops.  Massive propagators are represented by solid lines, while massless propagators appear as dashed lines. Following the convention in Table~\ref{tab:topologies2loopTT}, we have families {\tt topoTT1} and {\tt topoTT2} on the left and right, respectively. }
\end{figure}

  \exbox{
\inline{7}{& {\tt PM2n = AmplitudesMap[PM2, Times[\#,(nf-1)]\&, \{8, 16\}];} } 
}
The argument {\tt \{8, 16\}} specifies the amplitudes in {\tt PM2} that correspond to the $8^{\textrm{th}}$ diagram with the two projectors. The amplitudes in {\tt PM2n} can be processed by the function  {\tt AmplitudeToTopologies}. This requires the definition of the topologies. Since the goal is to reproduce the existent results in the literature, the topologies are defined as in ref. \cite{Fleischer1999}. For this, we observe that at two loops every self-energy diagram can be embedded into the kite integral family. Thus, it is enough to define our integrals according to the topologies in Table~\ref{tab:topologies2loopTT}. Their corresponding graphs are shown in fig. \ref{fig:T_topos}. 
The topologies are then defined in \ant~as 
\exbox{
\inline{8}{& {\tt DefineTopology[topoTT1,\{k1,k2\},\{p1\},\{Den[k1,0],Den[k2,MT],} \\
& {\tt \qquad\qquad\qquad\qquad\quad Den[k1-p1, MT],Den[k2-p1,0],Den[k1-k2,MT]\}]; } \\
}
\inline{9}{& {\tt DefineTopology[topoTT2,\{k1,k2\},\{p1\},\{Den[k1,MT],Den[k2,MT],} \\
& {\tt \qquad\qquad\qquad\qquad\qquad\, Den[k1-p1,0],Den[k2-p1,0],Den[k1-k2,0]\}]; }
}
}

\begingroup
\renewcommand*{\arraystretch}{1.3}
\begin{table}
    \centering
    \begin{tabular}{|w{c}{1cm}|w{c}{3.7cm}|w{c}{3.7cm}|}
       \hline 
       $D_i$  &  {\tt topoTT1} & {\tt topoTT2}   \\
       \hline
       \hline
       $D_1$  & $k_1^2$             & $k_1^2- m_t^2$    \\
       $D_2$  & $k_2^2 - m_t^2$     & $k_2^2 - m_t^2$   \\
       $D_3$  & $(k_1-p_1)^2-m_t^2$ & $(k_1-p_1)^2$     \\
       $D_4$  & $(k_2-p_1)^2$       & $(k_2-p_1)^2$     \\
       $D_5$  & $(k1-k_2)^2-m_t^2$  & $(k1-k_2)^2$      \\
       \hline
    \end{tabular}
    \caption{Definition of the two-loop topologies from ref. \cite{Fleischer1999} required for the computation of the top-quark self-energy in the SM.  }
    \label{tab:topologies2loopTT}
\end{table}
\endgroup

Alternatively, one can use the {\tt FindTopology} function of \ant\ to identify the topologies associated with the amplitude. Specifying the loop and external momenta and supplying the projector amplitude as input yields:
 \exbox{
 \inline{10}{& {\tt Topos2L = FindTopology[PM2n, DefineTopology -> False] }}\\
\breakline\\
 \outline{10}{& {\tt  \{\ TOPO1 -> \{Den[-k1, 0], Den[-k1 - k2, MT], Den[-k2, MT],}\\ 
   &{\tt Den[-k1 + p1, MT], Den[k2 + p1, 0]\},} \\
 &{\tt TOPO2 -> \{Den[-k1, 0], Den[-k1 - k2, 0], Den[-k2, 0],}\\
 &{\tt Den[-k1 + p1, MT], Den[k2 + p1, MT]\}\} 
}}}
We find that two distinct topologies suffice to describe all eight diagrams, one of which is the kite topology shown in fig.~\ref{fig:T_topos}. Because the kite topology already contains the five propagators needed to span all scalar products of the amplitude, no auxiliary propagators are required. 

In what follows, we perform the integral reduction using {\tt topoTT1} and {\tt topoTT2}.
Hence with these toplogies, we can then write the amplitude in terms of the topologies by issuing
\exbox{
\inline{12}{&{\tt FM2=AmplitudeToTopologies[PM2n,Topologies->\{topoTT1,topoTT2\}];}}
  }
  
The results in ref. \cite{Fleischer1999} are expressed in terms of eight master integrals defined in terms of only the topology {\tt topoTT1}. In our notation these integrals are 
\exbox{
\inline{13}{& {\tt basis=\{TopInt[topoTT1,1,1,1,1,0],TopInt[topoTT1,1,1,1,0,0],} \\
 & {\tt \qquad\; TopInt[topoTT1,1,1,1,1,1],TopInt[topoTT1,0,1,1,0,1],} \\
 & {\tt \qquad\; TopInt[topoTT1,0,1,1,0,2],TopInt[topoTT1,1,0,0,1,1],} \\
 & {\tt \qquad\; TopInt[topoTT1,1,0,0,1,2],TopInt[topoTT1,0,1,1,0,0]\};}  }
  }
We observe that, although there are two topologies, all the master integrals are derived from the first topology.

\exbox{
\inline{14}{&{\tt RM2=IntegralReduceKira[FM2, BasisMI -> basis];} }
 }
The list {\tt MasterIntegrals[]} contains the master integrals appearing in the amplitude after the reduction performed by \kira, by doing this we can confirm that the amplitude is given in terms of the preferred basis.

\section{Conclusion}
\label{sec:conclusion}
In this paper, we have 
introduced \ant, a \mathematica-based framework for computing scattering amplitudes in quantum field theory. The package enables the generation of Feynman diagrams via \qgraf~ for any QFT model, provided the model is already available with the package or its Feynman rules are provided in the prescribed \form~ format. 
The user can then compute squared amplitudes. Alternatively, he or she can extract scalar form factors from the Feynman diagrams using suitable projectors. The package
 includes a library of projectors for commonly used fields. The user may also define custom projectors to compute projected amplitudes tailored to their specific needs.

Once amplitudes are obtained, the package automates topology or integral family identification within \mathematica\ and a reduction to master integrals may be performed using dedicated interfaces to \litered\ or \kira. We illustrated the  capabilities of the package through examples of higher-order computations, demonstrating the use of its various functions. The goal of \ant~  is to provide a unified platform in \mathematica\ that integrates widely used tools in higher-order calculations, streamlining all stages of scattering amplitude computation.

\section*{Acknowledgments}
The authors thank Giuseppe Ventura, Ekta Chaubey and Yashavee Goel  for experimenting with the code and providing comments that helped to identify and fix bugs in the package. P.M\ also thanks Levente Fekésházy for providing valuable comments that helped improve the package.
The work of P.M\ was supported by the ERC Advanced Grant 101095857 Conformal-EIC. The work of C.D and A.V is supported by the DFG project 499573813 “EFTools”. Views and opinions expressed are however those of the author(s) only and do not necessarily reflect those of the European Union or the European Research Council. Neither the European Union nor the granting authority can be held responsible for them. 

\appendix
\section{Models}
\label{sec:models}
% !TEX root = anatar.tex

In this appendix, we describe the structure of the models files used by \ant. In principle, the user can overlook this section as long as there is no need to create a new model or modify existing ones. 

Currently, only the QED, SM and a variety of models with subsets of operators of the SMEFT are fully supported in \ant. For the latter, the chromomagnetic, modified Yukawa and gluon-fusion effective operators are included (see Table~\ref{tab:models}).

The model folder contains the following files:
\begin{itemize}
    \item {\tt General.m} contains general information about the fields, the gauge groups, and the parameters of the model.
    \item {\tt modelQG.qgraf} contains the information about the model required by \qgraf.
    \item {\tt Vertices.frm} contains the expressions for the vertices, written in terms of generic couplings.
    \item {\tt Couplings.frm} contains the definitions of the generic couplings in terms of the parameters of the model.
    \item {\tt Propagators.frm} contains the propagators for fields appearing as internal edges in the Feynman diagrams.
    \item {\tt Polarizations.frm} contains the polarizations and spinors assigned to the fields appearing as external edges in the Feynman diagrams.
\end{itemize}

In general, there are three types of files, which reflect the three steps that \ant~takes to output the amplitudes. The  file with the {\tt .m} extension is required for the processing of the input information given by the user in the \mathematica~interface. The {\tt .qgraf} file contains the required information for the generation of the Feynman diagrams in \qgraf. Finally, the files with the extension {\tt .frm} are files written in \form~language and contain in the explicit expressions for the objects that compose the amplitude. The diagram and amplitude generation is not done by \mathematica, but carried out in underlying runs.

The \qgraf~ output must be generated using the previously introduced {\tt array\_FR.sty} file. This writes every \qgraf~ index corresponding to the dual field of the propagators. The output of \qgraf~is then given in terms of three type of objects, namely {\tt pol(), pro(), vrtx()}, corresponding to the polarizations, propagators and vertices of fields, respectively. The arguments of such objects are fields, which in turn have as arguments a number tag, which provides the location of the field in the graph, and the momentum of the field. External fields are designated with a negative tag, while internal fields appear with positive tags. Thus, examples of such objects are {\tt pol(t(-1,p1))}, {\tt pro(G(1,2,k1-p1))} and {\tt vrtx(tbar(4,-k1),t(-1,p1),G(1,k1-p1))}, corresponding to the polarization of an external top quark, the gluon propagator and the $t\bar{t}G$ vertex, respectively. In what follows, we will refer to field objects in the \qgraf~output as {\tt F(a,pi)}. Let us notice that external momenta are always denoted by quantities like $p_i$, while loop momenta are given by $k_i$.

The objects given in the \qgraf~output do not carry information about the quantum numbers or masses of the fields, which are important for the Feynman rules in a given model. Thus, in the {\tt General.m} file all the information about the fields should be provided. This file contains lists clarifying the names for the fields and the anti-fields and if they are fermions or bosons. Additionally, other lists classify the fields depending on their spin, if they carry gluon or colour indices, and their mass in the case they are massive.  Finally, other lists are required providing the names of the structure constans and generators, as well as parameters of the model. The lists and variables required in the {\tt General.m} are presented in Table \ref{tab:generalFile}. The correct classification of the names of the relevant objects in a given model is required for their correct definition in \form. Unlisted and thus undefined objects will lead to errors, since every object must be declared in \form.

The classification of the fields discussed above allows \ant~to create for each list a corresponding {\tt set} in \form~language. These sets are used by \ant~to provide the quantum numbers that are not given in the output of \qgraf. This task is done recursively. In the first stage, using the {\tt Vertices.frm} file for the Feynman rules of the vertices, \ant~will provide a Lorentz index to a given field object according to the {\tt a} tag, such that for external fields the Lorentz index is of the type {\tt Lor1}, {\tt Lor2}, ...  In general, internal indices will have the prefix {\tt i} to any index type, such that we could have {\tt iLor}, {\tt iSpin}, {\tt iColour} and {\tt iGluon} type of indices. The Lorentz indices are added to all the fields in the model, independent on their quantum numbers, while the spin, gluon and colour indices are added in this order according to the sets obtained previously. For example, for the Higgs boson, the top quark, the gluon and the photon at the stage of the vertices we have the field objects {\tt H(-1,p1,Lor1)}, {\tt t(-2,p2,Lor2,Spin2,Colour2)}, {\tt G(-3,p3,Lor3,Gluon3)} and {\tt A(-4,p4,Lor4)}, respectively. It is important to highlight that the numbering of the tag {\tt a} and the momentum {\tt pi} might not correspond in the output of \qgraf, which means that it is possible to have field objects like {\tt G(-4,p3,Lor4,Gluon4)}. Given this index information, the vertices are given in terms of wildcard variables for pattern matching in \form, which means that such variables are indicated by attaching a question mark to the name of the variables themselves. A typical vertex is then provided as 
\newline

{\tt id vrtx(H(?a,p1?,Lor1?),H(?b,p2?,Lor2?),H(?c,p3?,Lor3?)=GC71*im;}
\newline 

The vertices, and in general any Feynman rules, are provided as an identity statement ({\tt id}) in \form. The symbol {\tt im} indicates the imaginary unit and the {\tt GC34} contains the constant factor in the vertex given in terms of the parameters of the model. For the moment we do not specify the explicit form of such {\tt GC} couplings. 

\begin{table}
\bgfb
\multicolumn{2}{c}{\textbf{Table~\ref{tab:generalFile}: Required variables of the {\tt General.m} } }\\
\\
\tt{{\tt AN\$ModelName }} & Name of the model given as a string.
\\
\tt{{\tt AN\$Fields }} & List of the names of the fields of the model.
\\
{\tt AN\$antiFields } & List of the names of the anti-fields of the model.
\\
{\tt AN\$Fermions } & List of the fermions. The fields in this list should carry spinor indices, thus ghost fields are not contained in this list, even though they are anti-commuting.
\\
{\tt AN\$Bosons } & List of bosons of the model. Anti-fields should be included in this list if they correspond to bosons.
\\
{\tt AN\$VectorFields } & List of vector fields. Anti-fields should be included in this list if they correspond to vector fields.
\\
{\tt AN\$ScalarFields } & List of scalar fields. Anti-fields should be included in this list if they correspond to scalar fields. Ghost and anti-ghost fields should also be included in this list. 
\\
{\tt AN\$xFermions } &  List of fermion fields.
\\
{\tt AN\$yFermions } &  List of anti-fermion fields.
\\
{\tt AN\$NoCouplings } & Variable designated for the number of Generic Couplings appearing in the file {\tt Couplings.frm}.
\\
{\tt AN\$Parameters } &  List of parameters, which should include the masses.
\\
{\tt AN\$Masses } & List containing tuples of the massive field and its corresponding mass, i.e., of the form {\tt \{\{H,MH\},\{Z,MZ\},\dots\}}.
\\
{\tt AN\$IndicesInFR } & Types of indices corresponding to unbroken gauge symmetries. For the SM this is {\tt \{Gluon,Colour\}}.
\\
{\tt AN\$FieldsClassified } & List specifying the fields that carry indices corresponding to gauge symmetries.
\\
{\tt AN\$StructureConstantsList } & List of structure constants of the model. 
\\
{\tt AN\$GeneratorsList } & List of generators of the model. 
\\
{\tt AN\$ExternalParameters } &  List containing the numerical values of the external parameters. This list is optional.
\egfb
\phantomcaption{\label{tab:generalFile}}
\end{table}

The second stage of inserting the Feynman rules is to substitute the polarizations of the fields. At this point, no indices need to be added to the field objects. In the file {\tt Polarizations.frm} all the identities for the polarizations can be found. The substitutions are based on the objects that will appear in the amplitudes as described in the next section (see Table~\ref{tab:objectsAmp}). As an example, let us look at the Feynman rule for the polarization of an initial tau lepton ({\tt ta}):
\newline

{\tt id pol(ta(?a,p1?initialMomenta,Lor1?,Spin1?))=SpinorU(?a,p1,MTA,Spin1);}
\newline

We notice that \ant~identifies if it is an initial or final fermion according to the momentum. Sets are defined for the initial and final momentum  from the declaration of the process. 

The final stage in the substitution of the Feynman rules corresponds to the propagators. Since the fields inside the propagator object produced in the output of \qgraf~contain the tag corresponding to the dual field, i.e., objects of the type {\tt pro(t(3,4,k1))}, it is required to add the respective {\tt Lor}, {\tt Spin}, {\tt Colour} and {\tt Gluon} dual indices. This is carried out in the file {\tt Propagators.frm}, in which a typical identity for the substitution of a given propagator looks like
\newline

{\tt id pro(G(?a,p1?,Lor1?,Lor2?,Gluon1?,Gluon2?)) }\\
{\tt \phantom{a}~~~~~~~~~~~~~~~~~~~~~~~~~=-im*Den(p1,0)*d\_(Gluon1,Gluon2)*(d\_(Lor1,Lor2));}
\newline

The symbol {\tt d\_(i,j)} is used for Kronecker delta and the metric with no distinction of the type of indices. Additionally, in the {\tt Propagators.frm} file we can find several procedures for different gauges for each sector of the model.

To work with more compact expressions \ant~works with generic couplings for the vertices. This speeds up the handling of large expressions in \form. The explicit expressions of the generic couplings can be found in the file {\tt Couplings.frm}. In the next paragraph we discuss in more detail how to substitute the explicit expressions of such couplings.

The Feynman rules in the model files must all be given in \form~language. The syntax in \form~language used in the Feynman rules is basically the same as the one used for the objects that compose the final expressions of the amplitudes in \mathematica~language, with minor changes and some exceptions. Most of the functions have the same header, such that the only change is that the round brackets {\tt (...)} in \form~become square brackets {\tt [...]} in \mathematica. Thus, the syntax for the objects relevant for the composition of Feynman rules can be found in the next section in Table \ref{tab:objectsAmp}. The exceptions to this are a few symbols and functions which in \form~are named with an underscore, for example the imaginary unit {\tt i\_}. Such syntax is not compatible with \mathematica. In Table \ref{tab:formSyntax} we list the exceptions that must be taken into account when writing Feynman rules in the model files. Let us notice that \ant~does not use the built in syntax of some objects in \form, as it is the case of the Dirac matrices, for which, instead of using {\tt g\_}, the syntax {\tt GammaM} is used.  
Similarly, for the $D$-dimensional Levi-civita tensor, the syntax {\tt Levieps} is used instead of the built-in 4-dimensional definition  {\tt e\_}. 

Finally, vertices which require dummy indices lead to some subtleties. In the SM, the only vertex of this type is the four-gluon interaction, where indices of the adjoint representation are summed in the products of structure constants. Such type of vertices appear more often in EFT's, where more complicated structures are taken into account. Each vertex that involves dummy indices should be wrapped in a {\tt repeat} statement where the executable statement is the vertex followed by a {\tt sum} statement of all the dummy indices.

\begin{table}
\bgfb
\multicolumn{2}{c}{\textbf{Table~\ref{tab:formSyntax}: Syntax of special symbols for Feynman rules. }}\\
\\
\tt{{\tt im}} & Imaginary unit.
\\
{\tt d\_(i,j)} & Metric tensor and Kronecker delta.
\\
{\tt sqrt\_(x)} & Square root of x.
\\
{\tt Levieps(Lor1,Lor2,Lor3,Lor4)} & $D$-dimensional Levi-civita tensor.
\egfb
\phantomcaption{\label{tab:formSyntax}}
\end{table}

\section{Model generation for \ant}
\label{sec:modelGen}

A $\beta$-version of an interface with \feynrules~for the generation of models admitted by \ant~can be found in the {\tt git} repository at
\begin{center}
\href{https://github.com/ANATAR-hep/FeynRules-Model-Interface}{https://github.com/ANATAR-hep/FeynRules-Model-Interface}
\end{center}
The interface can be obtained by cloning the repository using {\tt git}. This can be achieved by issuing the following command in any shell that has the git command available:
\begin{verbatim}
    https://github.com/ANATAR-hep/FeynRules-Model-Interface.git
\end{verbatim}
This module can be called by importing the file {\tt FOMain.m}. However, it depends entirely on \feynrules. Hence, for its correct implementation \feynrules~must be loaded in the usual manner. At the current stage, the module only works for BSM extensions with the same gauge group of the SM.\footnote{Models with four-fermions are not directly supported unless they are written using the auxiliary-mediator method.} It includes a single command to generate the model files
\newline\newline
{\tt WriteFFO[out,FeynRules,}\emph{options}{\tt ]}
\newline\newline
with {\tt out} the name of the output model and {\tt FeynRules} the Feynman rules obained in \feynrules~via the function {\tt FeynmanRules[]} (see the \feynrules~documentation \cite{Alloul:2013bka}). This command then produces a folder with all the model files. This output format, which we call \emph{FORM Format Output} (FFO), follows the guidelines presented in the Appendix \ref{sec:models}. The \emph{options} above stands for the set of options {\tt QGCouplings}, {\tt DeleteByFields} and {\tt SelectByFields}. The option {\tt QGCouplings} allows the user to provide a list of parameters that specify the coupling orders of the vertices in the \qgraf~model. Vertices that contain any insertion of some given fields can be ignored in the model generation routine by using the option {\tt DeleteByFields -> \{phi1,phi2,...\}}. Similarly, the option {\tt SelectByFields} keeps only those vertices with the specified fields.

Let us highlight that the input {\tt FeynRules} only stands for the Feynman rules, meaning that Lagrangians are not valid entries. Furthermore, the function {\tt WriteFFO[]} does not have any supported option that allows modifications of the Feynman rules. Hence, any particular substitution or modification that the user wants to perform on the Feynman rules must be done on the vertices before being provided to the function {\tt WriteFFO[]}. Finally, it is not possible to select the gauge for writing the propagators in the output model. Thus, by default, the produced {\tt Propagators.frm} file will always contain the Feynman rules for the propagators in both the Feynman and covariant gauges for the QCD sector, and the Feynman and unitary gauges for the EW sector.

As an example of the implementation of the interface, let us generate the corresponding \ant~model for the SM. We start by making sure that \feynrules~is loaded and the variable {\tt \$FeynRulesPath} properly set to the location where the file {\tt FeynRulesPackage.m} is found. Subsequently, we proceed as usual in \feynrules~and compute the Feynman rules by issuing
\exbox{
\inline{1}{& {\tt LoadModel["sm.fr"];}}\\
\inline{2}{& {\tt LoadRestriction["Massless.rst", "DiagonalCKM.rst"];}}\\
\inline{3}{& {\tt FRSM = FeynmanRules[LSM, FlavorExpand -> True];}}
}
Afterwards the module can be loaded as
\exbox{
\inline{4}{& {\tt \$FFOPath=SetDirectory[<path>];}}\\
\inline{5}{& {\tt <<"FOMain.m"}}\\
} 
Finally, to generate the model it is enough to do
\exbox{
\inline{6}{& {\tt WriteFFO["sm", FRSM, QGCouplings -> \{gs, ee, MT\}]}}
}
where we have chosen that the \qgraf~model file will have the  the couplings orders of vertices written in terms the strong coupling, the electron electric charge, and mass of the top, e.g., $g_s$, $e$ and $m_t$, respectively.
\section{On the $\gamma_5$-scheme for loop amplitudes }
\label{sec:gamma5}

In this appendix we comment on the $\gamma_5$-scheme for loop amplitudes scheme used when generating amplitudes in dimensional regularisation.
Any multi-loop calculation involving axial couplings in dimensional regularization (DR) encounters the challenge of properly defining the inherently four-dimensional objects, the Dirac $\gamma_5$ matrix and the Levi-Civita $\epsilon_{\alpha\beta\rho\sigma}$ symbol, in $D$ dimensions. In \ant, we have incorporated  the $D$-dimensional definition of $\gamma_5$  which was introduced by ’t Hooft-Veltman \cite{THOOFT1972189} and Breitenlohner-Maison \cite{Breitenlohner:1977hr} in dimensional regularisation.  
\begin{align}
\label{eq:gamma5}
        \gamma_5=\frac{i}{4!}\varepsilon_{\mu\nu\rho\sigma}\gamma^{\mu}\gamma^{\nu}\gamma^{\rho}\gamma^{\sigma}\,.
\end{align}
Using the convention $\varepsilon_{0123}=+1$ for the Levi-Civita symbol. However, the $\gamma_5$ defined through the above equation no longer fully anti-commutes with the $D$-dimensional $\gamma^{\mu}$, which has profound consequences in computations involving axial currents in $D$ dimensions.  In order to define the Hermitian axial current correctly we need to symmetrise it~\cite{Akyeampong:1973xi,Larin:1991tj} before using the definition \eqref{eq:gamma5}
\begin{align}
\label{eq:axial-symm}
        \gamma_{\mu}\gamma_5 \rightarrow \frac{1}{2} \Big(\gamma_{\mu}\gamma_5-\gamma_5\gamma_{\mu}\Big)\,.
\end{align}
By combining eqs.~\eqref{eq:gamma5} and \eqref{eq:axial-symm}, we obtain~\cite{Akyeampong:1973xi,Larin:1991tj}
\begin{align}
\label{eq:gamma5-axial}
        \gamma_{\mu}\gamma_5 = \frac{i}{6} \varepsilon_{\mu\nu\rho\sigma} \gamma^{\nu} \gamma^{\rho} \gamma^{\sigma}\,,
\end{align}
which is used in $D$ dimensions in \anatar. The contraction of pairs of Levi-Civita symbols appearing in the calculation is made through
\begin{align}
\label{eq:epsilon-contract}
        \varepsilon_{\mu_1\mu_2\mu_3\mu_4} \varepsilon_{\nu_1\nu_2\nu_3\nu_4}=
+~\mathrm{det}\begin{pmatrix} 
g_{\mu_1\nu_1}&g_{\mu_1\nu_2}&g_{\mu_1\nu_3}&g_{\mu_1\nu_4} \\ g_{\mu_2\nu_1}&g_{\mu_2\nu_2}&g_{\mu_2\nu_3}&g_{\mu_2\nu_4}\\
g_{\mu_3\nu_1}&g_{\mu_3\nu_2}&g_{\mu_3\nu_3}&g_{\mu_3\nu_4}\\
g_{\mu_4\nu_1}&g_{\mu_4\nu_2}&g_{\mu_4\nu_3}&g_{\mu_4\nu_4}
\end{pmatrix}   \,, 
\end{align}
where all the indices carried by space-time metric tensors on the right-hand side are (by definition) considered in $D$ dimensions.

%%%%%%%%%%%%%%%%%%%%%
% References 
%%%%%%%%%%%%%%%%%%%%%
\bibliographystyle{JHEP}
\bibliography{anatar}

\end{document}